%% file: main.tex
\documentclass[10pt,journal,compsoc]{IEEEtran}

\ifCLASSOPTIONcompsoc
  \usepackage[nocompress]{cite}
\else
  \usepackage{cite}
\fi
\usepackage{xspace}
\usepackage[utf8]{inputenc}
\usepackage{listings} 
\usepackage{booktabs} 
\usepackage[inline]{enumitem} 
\usepackage{adjustbox} 
\usepackage{hyperref} 
\usepackage{float} 
\usepackage{graphicx}
\usepackage{tcolorbox}
\usepackage[linesnumbered]{algorithm2e}
\usepackage{amsthm}
\usepackage{amsmath}
\usepackage{amssymb}
\usepackage{tabularx}
\usepackage[colorinlistoftodos,prependcaption,textsize=tiny]{todonotes}

\usepackage{tikz}
\usetikzlibrary{shapes.geometric, arrows.meta}
\tikzstyle{arrow} = [thick,->,>=stealth]
\tikzstyle{startstop} = [rectangle, rounded corners, minimum width=1.5cm, minimum height=.7cm,text centered, draw=black]
\tikzstyle{io} = [trapezium, trapezium left angle=70, trapezium right angle=110, minimum width=1.5cm, minimum height=.7cm, text centered, draw=black]
\tikzstyle{process} = [rectangle, minimum width=1.5cm, minimum height=.7cm, text centered, draw=black]
\tikzstyle{decision} = [diamond, minimum width=1.5cm, minimum height=.7cm, text centered, draw=black]

\newtheorem{definition}{Definition}

\usepackage{colortbl}
\definecolor{pblue}{rgb}{0.13,0.13,1}
\definecolor{pgreen}{rgb}{0,0.5,0}
\definecolor{pred}{rgb}{0.9,0,0}
\definecolor{pgrey}{rgb}{0.46,0.45,0.48}

\newcommand{\pit}{PITest\xspace}
\newcommand{\descartes}{Descartes\xspace}

\newcommand{\ie}{\textit{i.e.}\xspace}

\newcommand{\etal}{\textit{et al}\xspace}
\newcommand{\java}{Java\xspace}
\newcommand{\junit}{JUnit\xspace}
\newcommand{\github}{Gihub\xspace}
\newcommand{\maven}{Maven\xspace}
\newcommand{\evosuite}{\textsc{Evosuite}\xspace}
\newcommand{\dspot}{DSpot\xspace}

\newcommand{\ourpreviouspaper}{\cite{vera_comprehensive_2018}}

\newcommand{\tool}{\textsc{Reneri}\xspace}

\usepackage[framemethod=tikz]{mdframed}
\mdfdefinestyle{mpdframe}{
       frametitlebackgroundcolor   =black!15,
       frametitlerule              =true,
       roundcorner                 =5pt,
       middlelinewidth             =0.8pt,
       innermargin                 =0.2cm,
       outermargin                 =0.2cm,
       innerleftmargin             =0.2cm,
       innerrightmargin            =0.2cm,
       innertopmargin              =0.2cm,
       innerbottommargin           =0.2cm
}

\newcommand{\takeaway}[2]{
    \begin{mdframed}[style=mpdframe]Answer to \ref{#1}: #2\end{mdframed}
}

\begin{document}

\title{Suggestions on Test Suite Improvements with Automatic Infection and Propagation Analysis}

\author{Oscar~Luis~Vera-P\'erez, 
        Benjamin~Danglot,
        Martin~Monperrus,
        and~Benoit~Baudry
\IEEEcompsocitemizethanks{\IEEEcompsocthanksitem O. Vera-P\'erez is with Univ Rennes, Inria, CNRS, IRISA ,France
\IEEEcompsocthanksitem B. Danglot is with Inria Lille - Nord Europe
\IEEEcompsocthanksitem M. Monperrus and B. Baudry are with KTH Royal Institute of Technology}%
\thanks{This work has been partially supported by the EU Project STAMP ICT-16-10 No.731529 and by the Wallenberg Autonomous Systems and Software Program (WASP) funded by the Knut and Alice Wallenberg Foundation.}}

\IEEEtitleabstractindextext{%
\input{abstract}
}

\maketitle

\input{content}

\bibliographystyle{IEEEtran}
\bibliography{references}


\end{document}

%% file: abstract.tex
\begin{abstract}
    An extreme transformation removes the body of a method that is reached by one test case at least. If the test suite  passes on the original program and still passes after the extreme transformation, the  transformation is said to be undetected, and the test suite needs to be improved.
    In this work we propose a technique to automatically determine which of the following three reasons prevent the detection of the extreme transformation is :  the test inputs are not sufficient to \emph{infect} the state of the program;  the infection does not \emph{propagate} to the test cases;  the test cases have a \emph{weak oracle} that does not observe the infection.
    We have developed \tool, a tool that observes the program under test and  the test suite in order to determine runtime differences between test runs on the original and the transformed method. The observations gathered during the analysis are processed by \tool to suggest possible improvements to the developers. 
    
    We evaluate \tool on 15 projects and a total of  312 undetected extreme transformations. The tool is able to generate a suggestion for each each undetected transformation. 
    For 63\% of the cases, the existing test cases can infect the program state, meaning that undetected transformations are mostly due to observability and wek oracle issues. Interviews with developers confirm the relevance of the suggested improvements and experiments with state of the art automatic test generation tools indicate that no tool can improve the existing test suites to fix all undetected transformations.
\end{abstract}

%% file: content.tex

\newcommand\tuet{312} 
\newcommand\tpi{198} 
\newcommand\tni{\the\numexpr\tuet-\tpi\relax} 
\newcommand\percentage[2]{\the\numexpr(#1)*100/(#2)\relax\%} 
\newcommand\percentoftotal[1]{\percentage{#1}\tuet}
\newcommand\tbool{129} 
\newcommand\tboolni{65} 
\newcommand\twor{83} 
\newcommand\tnp{115} 

\newcommand\attempts{ten} 

\newcommand\essol{146}   
\newcommand\dssol{180}   
\newcommand\esdsun{239}  
\newcommand\esdsin{87}   
\newcommand\esdsdiff{59} 
\newcommand\dsesdiff{93} 
\newcommand\unsolved{73} 

\newcommand\solni{81}
\newcommand\solnp{92}
\newcommand\solwo{66}

\section{Introduction}

Niedermayr \etal \cite{niedermayr_will_2016} define an extreme program transformation as one that removes all instructions in the body of a method. If needed, the body is replaced with a single return instruction producing a fixed constant value. If we apply an extreme transformation to a method, and no test case is able to detect this change, this is an indicator that the code of this method is not correctly assessed by the test suite.

Our previous work \ourpreviouspaper{} demonstrated two key phenomena:
\begin{enumerate*}
    \item the methods where an extreme transformation is not detected are indeed the least tested in the program, and
    \item undetected extreme transformations provide valuable information to the developers in order to improve their test suite.
\end{enumerate*}

Undetected extreme transformations provide valuable insights about which methods should be better tested. 
Developers can improve the test suite by enhancing an existing test case with more inputs or assertions, or they can add a new test. 
Yet, the complex interactions between the test cases and the program under test make it challenging to understand what action to take.
For example, the method to be tested can be executed only at the end of a long sequence of invocations, or it can be private and indirectly executed by the test suite. In all these cases, developers need to spend significant effort to understand the complex chain of invocations between the test cases and the method to be tested, its effect on the method and how the extreme transformation affects the program state.

In this work, we propose a tool, called \tool, which automates the  analysis of the test execution on both the original and the transformed method. With \tool, we aim at automating the analysis of the interaction between the test suite and the method to be tested in order to assist the developer in the improvement of the test suite.
\tool implements a dynamic analysis inspired by the \emph{Reachability, Infection, Propagation} model (\emph{RIP} or \emph{PIE}\footnote{Propagate Infect Execute})\cite{morell_theory_1990,demillo_constraint-based_1991,voas_pie:_1992}. The tool takes a program, its test suite and a set of undetected extreme transformations as inputs. \tool instruments the code of the program and the test suite to gather information about the program state during test executions. Then, it compares the information produced during the execution of the original code with the information from the execution of the extreme transformation.

\tool determines the root cause of the undetected transformation:
\begin{enumerate*} 
    \item the test inputs are not sufficient to \emph{infect} the state of the program (\textbf{no-infection});
    \item the infection does not \emph{propagate} to the test cases (\textbf{no-propagation});
    \item the test cases have a \emph{weak oracle} that does not observe the infection (\textbf{weak-oracle}). 
\end{enumerate*}

\tool uses the diagnosis information to synthesize suggestions, in plain English, that can be directly followed by the developers to create a new test input, add a new assertion or create a new test case.

We perform three systematic evaluations of \tool: 
\begin{enumerate*}
    \item a quantitative analysis of root causes behind undetected extreme transformations; 
    \item a qualitative analysis of the synthesized suggestions through interviews with developers; 
    \item an quantitative evaluation of test generation tools to handle the undetected transformations. 
\end{enumerate*}

The first evaluation is performed with 15 projects containing 312 undetected transformations. We observe that there is no dominant root cause to miss a transformation, there is balance between \textbf{no-infection}, \textbf{no-propagation} and \textbf{weak oracle}.

In the second evaluation, we use \tool to generate improvement suggestions for a selected number of undetected extreme transformations in four open-source projects. Developers considered these suggestions helpful in most cases and also provided feedback on \tool's strengths and weaknesses.

In the third evaluation, we design and implement two strategies to handle the undetected transformations: one based on \evosuite as a test generation alternative and another based using \dspot as a test case improvement approach. In total, both strategies were able to generate test cases that detect \percentage\esdsun\tuet\ of the previously undetected extreme transformations.

The key contributions of this work are as follow:
\begin{itemize}
\item
	\item \tool, a new tool that automates the infection and propagation analysis on \java{} programs. The key novelty is that it generates natural language suggestions that can be followed by the developers to improve their test suite
	\item an empirical assessment of the tool with 15 projects and a total of 312 undetected extreme transformations. This analysis indicates that undetected extreme transformations are mostly due to a lack of observation.
	\item open science: \tool is available as open source \footnote{https://github.com/STAMP-project/reneri} and empirical data is provided as open data \footnote{https://github.com/STAMP-project/descartes-amplification-experiments} to facilitate future replication
\end{itemize}

The rest of this paper is organized as follows: first we describe extreme transformations through an example and discuss how they can be helpful to reveal weaknesses in the test suite; then we define the main concepts behind the dynamic analysis implemented in our tool and how we adapt the \emph{RIP} model conditions to extreme transformations. In the following sections we describe the different stages of the proposed dynamic analysis and how they are implemented. Then, we expose the research questions we followed to validate our proposal, and the answers to these question. Finally we discuss the threats to the validity of our study, a list of related works and the conclusions that can be drawn from our present work.

\section{Extreme transformations and test suite weaknesses}

In this section we go through an example that illustrates the possible interactions between a test suite, a program under test and extreme transformations. We show what types of weaknesses in the test suite these extreme transformations can reveal. We also illustrate how tailored dynamic analyses can be used to guide the developers for test suite improvement. 

\subsection{Examples of undetected extreme transformations}

\autoref{list:example:class} displays the \texttt{VersionedSet} class. It implements a set that keeps track of the insertion operations through the \texttt{version} field. 
\autoref{list:example:test-suite} shows a \junit{} test class that verifies \texttt{VersionedSet}. The test class contains three test methods implementing three test cases. These test cases are able to reach all methods in the class under test, that is, at least one statement of each method is executed by a test case. All tests pass when executed with the original class code. 

For this example, the extreme transformation process \ourpreviouspaper{} analyzes the \texttt{VersionedSet} class and the associated test class and performs seven operations: 
\begin{enumerate*} 
	\label{example-transformations} 
	\item \label{trans:example:add} remove the body of \texttt{add} (line \ref{line:example:add})
	\item \label{trans:example:version} remove the body of \texttt{incrementVersion} (line \ref{line:example:incrementVersion}) 
	\item \label{trans:example:equals-true} replace the body of \texttt{equals} by \texttt{return true} (line \ref{line:example:equals})
	\item \label{trans:example:equals-false} replace the body of \texttt{equals} by \texttt{return false} 
	\item \label{trans:example:intersect} replace the body of \texttt{intersect} by \texttt{return null} (line \ref{line:example:intersect})
	\item \label{trans:example:empty-true} replace the body of \texttt{isEmpty} by \texttt{return true} (line \ref{line:example:isEmpty})
	\item \label{trans:example:empty-false} replace the body of \texttt{isEmpty} by \texttt{return false}
\end{enumerate*}

 Methods \texttt{size}, \texttt{contains} and \texttt{getVersion} are ignored by the transformation process \ourpreviouspaper{}: \texttt{size} and \texttt{contains} delegate the execution to similar methods of the underlying instance of \texttt{ArrayList}. \texttt{getVersion} is a simple getter for the \texttt{version} field. Methods matching these patterns are automatically detected and skipped. This design decision relies on the observation that most developers do not target these types of methods directly in test cases.

\begin{lstlisting}[caption=An example Java class under test, captionpos=b, label=list:example:class]
public class VersionedSet {
 private long version = 0;
 private ArrayList elements = new ArrayList();

 public void add(Object item) { %*\label{line:example:add}*)
  if (elements.contains(item)) return;
  elements.add(item);
  incrementVersion();
 }

 private void incrementVersion() { version++; } %*\label{line:example:incrementVersion}*)

 protected long getVersion() { return version; } %*\label{line:example:getVersion}*)

 public int size() { return elements.size(); } %*\label{line:example:size}*)

 public boolean isEmpty() { return size() == 0; } %*\label{line:example:isEmpty}*)

 public boolean contains(Object item) { %*\label{line:example:contains}*)
  return elements.contains(item); 
 }

 @Override
 public boolean equals(Object otr) { %*\label{line:example:equals}*)
  if(!(otr instanceof VersionedSet))
   return false;
  VersionedSet otrSet = (VersionedSet)otr;
  if(otrSet.size() != size())
   return false;
  for (Object item : elements) {
   if(!otrSet.contains(item))
        return false;
  }
  return true;
 }

 public VersionedSet intersect(VersionedSet otr) { %*\label{line:example:intersect}*)
  if (isEmpty() || otr.isEmpty())
   return new VersionedSet();
  VersionedSet result = new VersionedSet();
  for(Object item : elements) {
   if(otr.contains(item))
    result.add(item);
  }
  result.version = 0;
  return result;
 }
}
\end{lstlisting}

\begin{lstlisting}[caption=Test class verifying \texttt{VersionedSet}, captionpos=b, label=list:example:test-suite]
public class VersionedSetTest {
 @Test
 public void testAdd() { %*\label{line:example:testAdd}*)
  VersionedSet list = new VersionedSet();
  list.add(1); %*\label{line:example:after-add}*)
  assertEquals(1, list.size()); %*\label{line:example:insert-assertion}*)
 }

 @Test
 public void testEquals() { %*\label{line:example:testEquals}*)
  VersionedSet one = new VersionedSet();
  VersionedSet two = new VersionedSet();
  assertTrue(one.equals(two)); 
 }

 @Test
 public void testIntersection() { %*\label{line:example:testIntersection}*)
  VersionedSet one = new VersionedSet();
  one.add(1);
  VersionedSet two = new VersionedSet();
  two.add(2);
  VersionedSet result = one.intersect(two); %*\label{line:example:observed-propagation}*)
  assertFalse(result.contains(1));
  assertFalse(result.contains(2));
 }
}
\end{lstlisting}

The objective of the extreme transformations is to run the test class against each transformed method. This provides an assessment of the test suite ability to reveal these transformations. For our example, we observe the following: transformations \ref{trans:example:add}, \ref{trans:example:equals-false} and \ref{trans:example:intersect} are detected: at least one of the test cases fails when the transformation is performed. Meanwhile transformations \ref{trans:example:version}, \ref{trans:example:equals-true}, \ref{trans:example:empty-true} and \ref{trans:example:empty-false} are not detected by the test suite.

\subsection{Conditions to detect extreme transformations}
\label{sec:conditions}

In order for the test suite to detect an extreme transformation, the following conditions must hold (we adapt these conditions from the \emph{Reachability, Infection, Propagation} model): 
\begin{enumerate*}
	\item \textbf{reach:} at least one test case invokes the method modified by the extreme transformation (reaches the transformed method)
	\item \textbf{infect:} the program state after the method call must be different from the state observed when  no transformation is performed (the transformation infects the state)
	\item \textbf{expose:} the modification propagates to a point in the top level code of a test case (the infection propagates to an observable point)
	\item \textbf{assert:} there is an oracle (\ie{} an assertion) that verifies the modified state
\end{enumerate*}.

The \textbf{reach} condition depends on the coverage achieved by the test suite. In this paper we focus only in the analysis of methods that are reached by the test suite, that is, at least one statement in their code is executed. In our example, all methods meet this condition.

For the \textbf{infect} condition, we consider the following: after the execution of the transformed method, we observe the state of the return value, the state of the object instance on which the method has been invoked, and the state of any of the actual parameters. 
We compare the values when running the test suite on the original and the transformed versions of the method compared. If any value is different between these executions, we consider that the interaction between the test suite and the transformation can infect the program state.  

An infection is \textbf{exposed}  if it can be observed from a test case, \ie, if the test case can query some program states that are affected by the infection at the boundary of the transformed method. The last condition about \textbf{assert} holds if the test case actually checks a property that is influenced by this visible state that is affected by the infection.

\subsection{Analyzing undetected extreme transformations}

The analysis of the previous conditions helps to explain why an extreme transformation is not detected by the test suite. We illustrate this with our example.

The \texttt{equals} method (line \ref{line:example:equals}) is reached by \texttt{testEquals} (line \ref{line:example:testEquals}). Yet, the extreme transformation that replaces the body of \texttt{equals} by \texttt{return true} has no effect on the immediate state of the program, when executed with \texttt{testEquals}. The \texttt{equals} method returns the same value for the given input and has no other side effect. In fact, the test case makes the method return \texttt{true} and no other test case makes the method return \texttt{false}. The transformation is reached in the execution but the program state is not infected by the extreme transformation. The effects are therefore not propagated and the transformation can not be detected. In this particular example it is necessary to augment the test suite with a new test case using a new input.

When transformation \ref{trans:example:version} is applied the method \texttt{incrementVersion} (line \ref{line:example:incrementVersion}) and \texttt{testAdd} is executed, the immediate program state is infected: the value of \texttt{version} is 0 while it is 1 with the original method body. The infection is propagated to the code of the test case, as the state of the \texttt{list} variable is affected after the invocation of \texttt{add} in line \ref{line:example:after-add}. However, the test case is not verifying this state element. In this example it is necessary to add a new assertion targeting \texttt{getVersion}.

Both transformations \ref{trans:example:empty-false} and \ref{trans:example:empty-true} affect the method \texttt{isEmpty}. The former transformation makes the method return \texttt{false}. In the context of the \texttt{testIntersection} test case, this transformation is similar to \ref{trans:example:equals-true} for the \texttt{equals} method. The program state is not infected. As for the latter transformation, the program state is infected. \texttt{isEmpty} should return \texttt{false} and not \texttt{true}. However, the return value of \texttt{intersect} turns out to be the same as expected in the original test case and therefore the infection is not propagated. This analysis could produce two possible outcomes. The utility of \texttt{isEmpty} for the implementation of \texttt{intersect} could be questioned. No matter the result of this method, the final effect is the same so the code could require some refactoring. On the other hand, a completely new test case could handle these two transformations. In particular, a test case that can target directly the \texttt{isEmpty} method, as it is public.

These three examples illustrate how we can obtain valuable insights when analyzing the interactions between test cases and extreme transformations at the immediate program state after the transformation. Such an analysis can support the developers in deciding whether they should add assertions to an existing test case, or add a new test case or if the program is not fully testable and requires refactoring.

\section{Automatic Synthesis of Suggestions for Test Improvement}
\label{sec:technical-contribution}

The automatic synthesis of suggestions for test improvements takes as input a \java{} program, its test suite, and a set of undetected extreme transformations. These transformations can be discovered from the program and the test suite following the proposal described in our previous work \ourpreviouspaper{}. The approach consists in automatically understanding why a given extreme transformation is not detected and in suggesting the developers a possible test improvement to detect it. These understanding and suggestion synthesis activities are based on the comparison of the test suite execution on the original and the transformed program. 

In this section, we provide precise definitions of the program analysis concepts that we manipulate to understand the effect of extreme transformations. Then, we detail each step of the dynamic analysis and test improvement suggestion synthesis.

\subsection{Definitions}
\label{sec:definitions}

We now rigorously define our terminology for state observation. Our definitions are scoped by the \java{} language and the JVM. We believe they could be extrapolated to other languages and runtimes in general.
 
\begin{definition}
	\label{def:basic-state}
	\textbf{Basic state of a value:}
	Let $v$ be a \java{} typed value, the \textit{basic state} of $v$ is a set of property-value pairs $\mathit{BS}$ defined as follows:
	\begin{itemize}
		\item If the type of $v$ is any \java{} primitive type, any wrapper type or \texttt{String} , then $\mathit{BS} = {(\mathit{value}, v)}$.
		\item If $v$ is a reference to an object, then $\mathit{BS}$ contains $(\mathit{null}, b)$ where $b$ is a boolean value indicating whether $v$ is the \texttt{null} reference.
		\item If $v$ is a reference to an array, then $\mathit{BS}$ contains $(\mathit{length}, l)$ where $l$ is the length of the array.
		\item If the type of $v$ implements interfaces \texttt{java.util.Collection} or \texttt{java.util.Map}, then $\mathit{BS}$ contains $(\mathit{size}, l)$ where $s$ is the size of the collection.
	\end{itemize}
\end{definition}

\begin{definition}
	\label{def:state}
	\textbf{State of value:}
	The state of a given value $v$ is a set of properly-value pairs $S$ defined as follows:
	\begin{itemize}
		\item $S$ contains all the elements in the \textit{basic state} of $v$.
		\item If the type of $v$ is not a \java{} primitive type, a wrapper type or \texttt{String} then $S$ contains the \textit{basic state} of all values of all fields, declared or inherited by the type of $v$.
	\end{itemize}
\end{definition}

For example, the state of a reference to an instance of the \texttt{VersionedSet} class in \autoref{list:example:class} right after its creation would be: $\{ (\mathit{null}, \mathit{false})$ $(\mathit{version}, 0),$ $(\mathit{elements.null}, \mathit{false}),$ $(\mathit{elements.size}, 0)\}$. 

In this paper we consider a program $P$ to be a set of method and its accompanying test suite $T$ to be a set of test cases. 

\begin{table}
	\caption{Predefined values used to obtain the extreme transformations.}
	\label{tab:desops}
	\centering
	\begin{tabular}{ll}
		\toprule
		Method type         & Values used \\
		\midrule
		void                & -           \\
		Reference types     & null        \\
		boolean             & true,false  \\
		byte,short,int,long & 0,1         \\
		float,double        & 0.0,0.1     \\
		char                & ` ', `A'   \\
		String              & ``'', ``A''\\
		T[]                 & new T[]\{\} \\
		\bottomrule
	\end{tabular}
\end{table}

\begin{definition}
	\label{def:et}
	\textbf{Extreme transformation:}
	We consider an extreme transformation to be a tuple $(m,m')$ where $m \in P$, $m'$ is the transformed variant of $m$, also called transformed method. $m'$ is obtained from $m$ by removing all the instructions in its body and adding a single \textbf{return} instruction with a predefined value if $m$ returns a value. The predefined values according to the return type of $m$ are shown in \autoref{tab:desops}. 
	An extreme transformation is said to be \emph{undetected} if $\nexists t \in T$ that fails when $m$ is replaced by $m'$ in $P$ and the tests in $T$ are executed.
\end{definition}

Our analysis takes as input a program $P$, a test suite $T$ and $U$, a set of undetected extreme transformations as defined in \autoref{def:et}.

\begin{definition}
	\label{def:method-state}
	\textbf{Local immediate program state after a method invocation:}
	The \textit{local immediate state} of a program $P$ after the execution of a method $m \in P$ is a set of property-value pairs $\mathit{MS}_m$ defined as follows:
	\begin{itemize}
		\item $\mathit{MS}_m$ contains all elements in the state (as defined in \autoref{def:state}) of all the arguments of $m$, if any.
		\item If $m$ has a return value, (\ie{} $m$ is not \texttt{void}), then $\mathit{MS}_m$ contains all elements in the state of the result value as defined in \autoref{def:state}.
		\item If $m$ is not static, then $\mathit{MS}_m$ contains all elements in the state (as defined in \autoref{def:state}) of the instance on which $m$ was invoked.
	\end{itemize}
	$\mathit{MS}_{m,t}$ denotes the immediate program state after the invocation of $m$ when executing the test case $t \in T$. 
\end{definition}

As an example, if a program creates an instance of \texttt{VersionedSet} and then invokes \texttt{addElement} on this instance with a \texttt{String} \texttt{"something"}, the local program state would be: $\{ (\mathit{null}, \mathit{false})$ $(\mathit{version}, 1),$ $(\mathit{elements.null}, \mathit{false}),$ $(\mathit{elements.size}, 1),$ $(\mathit{value}, \mathit{"something"})\}$. Notice how this state includes the state of the $VersionedSet$ instance and the state of the argument. In this case the method is \texttt{void}, so the result value is not reflected in the state.

\begin{definition}
	\label{def:infection}
	\textbf{Local immediate program state infection:}
	Let $m \in P$ be a method that is reached (executed) by at least one test case $t \in T$. Let $(m,m') \in U$ be an undetected transformation. We say that the execution of the test case $t$ provokes an infection of the local immediate program state under the extreme transformation if $\mathit{MS}_{m,t} \neq \mathit{MS}_{m',t}$. That is, there is an infection if we are able to observe a difference in the immediate program state after the execution of $m$ by $t$ and the execution of $m'$ by the same test case $t$.
\end{definition}

\begin{definition}
	\label{def:test-state}
	\textbf{Test case state:}
	Let $t \in T$ be a test case. Let $O$ be the set of values local to $t$, that is, all the values that result from any expression or subexpression in the code of $t$. Then, $\mathit{TS}_t$ denotes the state of $t$ and it is the union of the states of all values local to $t$. That is $\cup_{o \in O}{S_o}$. 
	$\mathit{TS}_m$ denotes the state of $t$ when the original method $m \in P$ is used and $\mathit{TS}_{t,m'}$ denotes the state of $t$ when it is executed under the extreme transformation $(m, m')$.
\end{definition}

For example, the state of \texttt{testEquals} in line \ref{line:example:testEquals} of \autoref{list:example:test-suite}  includes the state of the set instances referenced by \texttt{one} and \texttt{two} and the result of the result of the invocation to the \texttt{equals} method.

\begin{definition}
	\label{def:propagation}
	\textbf{Program state infection propagation:}
	Let $m \in P$ be a method and $(m,m') \in U$ an undetected extreme transformation. The immediate program infection is said to propagate to the test case $t \in T$ if there is a difference between the state of $t$ while executing $m$ and the state of $t$ when executing $m'$. That is if $\mathit{TS}_m \neq \mathit{TS}_{t,m'}$.
\end{definition}

As per the definitions above, we study the local state of a program after a method invocation and the local state of a test case. The former includes the state of the receiver of the method, the state of the arguments, and the state of the resulting value. The latter includes the state of all values resulting from an expression in the code of the test case.

\subsection{Overview of the process for test improvement suggestions synthesis}
\label{sec:overview}

The global process to understand undetected extreme transformations and to synthesize a test improvement suggestion operates in three main stages. Algorithm \ref{alg:process} outlines the process. Each stage is detailed in the following sections:
\begin{enumerate}
	\item Infection detection: identify the extreme transformations that infect the local immediate state; (line \ref{line:algo:infection}, \autoref{sec:infection})
	\item Propagation detection: discover which infections reach some test case states and (line \ref{line:algo:propagation}, \autoref{sec:propagation})
	\item Test improvement suggestion: generate the report by consolidating the information gathered in the two previous stages (line \ref{line:algo:hints}, \autoref{sec:hints}). 
\end{enumerate}

The initial two stages include a dynamic analysis of the program. Each stage instruments the elements to be observed and executes the test suite with the original code and the transformed method variant. These executions of the instrumented program record the program and test states to help the suggestion synthesis. The states are compared to discover the symptom that explains each why each transformation was not detected by the test suite.

\begin{algorithm*}
	\DontPrintSemicolon
	\KwData{$P$: program under test, $T$: test suite, $U$: undetected extreme transformations}
	\SetKwFunction{Finfections}{find\_infections}
	\SetKwFunction{Fpropagations}{find\_propagations}
	\SetKwFunction{Fhints}{gen\_suggestions}
	
	\textit{no-infection}, \textit{infection} $\gets$ \Finfections{$P$, $T$, $U$} \; \label{line:algo:infection}
	\textit{no-propagation}, \textit{weak-oracle} $\gets$  \Fpropagations{$P$, $T$, \textit{infection}} \; \label{line:algo:propagation}
	\Fhints{\textit{no-infection}, \textit{no-propagation}, \textit{weak-oracle}} \; \label{line:algo:hints}
	\caption{The three stages process implemented by \tool}
	\label{alg:process}
\end{algorithm*}

The  analysis results in the identification of one symptom for each undetected extreme transformation. The identified symptom shall help to explain why the extreme transformation was not detected by the test suite. We identify the following three types of symptom:
\begin{itemize}
	\label{enum:hints}
	\item \textbf{no-infection} (\textbf{ni}):  there is no observable difference in the local state of the program after the method invocation when the test case is executed on the original and transformed method.
	\item \textbf{no-propagation}(\textbf{np}): there is an observable difference in the program state at the transformation point, but there is no observable difference in the state of the test case.
	\item \textbf{weak-oracle} (\textbf{wo}): the program state infection is propagated to the code of a test case but no assertion fails.
\end{itemize}

These symptoms, as well as intermediate observations of the dynamic program analysis are consolidated in a final report to provide concrete improvement suggestions to the developers.

\subsection{Infection detection}
\label{sec:infection}

\begin{algorithm}
    \DontPrintSemicolon
    \KwData{$P$: program under test, $T$: test suite, $U$: set of undetected extreme transformations}
    \KwResult{\textit{no-infection}: \{\textbf{no-infection} symptoms\}, \textit{infection}: \{extreme transformations  that produce an infection\}}
    \SetKwFunction{Finfections}{find\_infections}
    \SetKwProg{Fun}{function}{does}{end}
    \SetKwFunction{Fobserve}{observe}
    \SetKwFunction{Fgetdiff}{get\_diff}
  
    \Fun{\Finfections{$P$, $T$, $U$}}{
      \ForEach{$ (m,m') \in U $}
      {
        $T_m \gets $ \{ $t \in T \colon t$ executes $m$ \} \;
        $P_i \gets$ instrument $m$ in $P$ \; \label{line:algo:instrument}
        $\mathit{MS}_m \gets $ \Fobserve{$P_i$, $T_m$} \;
        $P' \gets$ replace $m$ by $m'$ in $P$ \;
        $P'_i \gets$ instrument $m'$ in $P'$ \; \label{line:algo:instrument2}
        $\mathit{MS}_{m'} \gets$ \Fobserve{$P'_i$, $T_m$} \;
        $\mathit{diff} \gets$ \Fgetdiff{$\mathit{MS}_m$, $\mathit{MS}_{m'}$} \;
        \eIf{ $ \mathit{diff} \neq \emptyset $ } {
          $\mathit{infection} \gets \mathit{infection} \cup \{ (m, m'), \mathit{diff}) \} $ \;
        }{
          \textit{no-infection} $\gets$ \textit{no-infection} $\cup \{ (m, m') \} $ \;
        }
      }
      \KwRet \textit{no-infection}, \textit{infection} \;
    }
    \caption{Infection detection stage of the process. This stage identifies the extreme transformations that are not detected because of a \textbf{no-infection} symptom}
    \label{alg:infection}
\end{algorithm}

\begin{algorithm}
    \DontPrintSemicolon
    
    \SetKwFunction{Fobserve}{observe}
    \SetKwFunction{Fgetdiff}{get\_diff}
    \SetKwProg{Fun}{function}{does}{end}
  
    \Fun{\Fobserve{$P$, $T$}}{
      \For{$i \gets 1$ to $N$}{
        $S_i \gets$ execute $T$ with $P$ \;
      }
      \Return{ $\cup^{N}_{i=1} S_i$ }
    }
  
    \bigskip
  
    \Fun{\Fgetdiff{$S_1$, $S_2$}}{
      $\mathit{diff} \gets \emptyset$ \;
      \ForEach{$(p,v) \in S_1$}{
        \If{ $ \exists (p,w) \in  S_2 \land v \neq w $ }{
          $\mathit{diff} \gets \mathit{diff} \cup \{(p,v,w)\}$ \;
        }
      }
      \Return{$\mathit{diff}$}
    }
    \caption{Two functions to compute the state invariants across test executions and find a difference between the state of the original program and the transformed variant.}
    \label{alg:diff}
\end{algorithm}

The goal of the first stage is to determine whether the execution of the tests under an extreme transformation infects the program state. We instrument the original program and the program after the application of the extreme transformation to observe the immediate program state in both situations. If we can observe a difference between the states, then this signals an infection. The main steps of this stage are outlined in \autoref{alg:infection}

\subsubsection{Instrumentation} 

For each method and its transformed variants we observe the following parts of the program state: the instance on which the method is invoked, all arguments (if any), and the result value (if the method returns a value). 

For primitive values and strings, the observation is trivial. For objects of other types, an observation means inspecting public and private fields using reflection. In this way the observation minimizes potential side-effects to the object being observed.

In practice, we obtain these observations by instrumenting the code of the original methods and its transformed variant (lines \ref{line:algo:instrument} and \ref{line:algo:instrument2} from \autoref{alg:infection}). The instrumentation targets the compiled bytecode of the program.

\autoref{list:instrumentation:infection} shows an example of how the \texttt{equals} method from \autoref{list:example:class} is instrumented for observation. The \texttt{observe} function saves the state of the object as defined in \autoref{def:state}.

\begin{lstlisting}[caption=Method instrumented to observe the  immediate program state, captionpos=b, label=list:instrumentation:infection]
public boolean equals(Object other) {
 boolean result = ... /* original method code */
 /* Observation */
 observe(this);
 observe(other);
 observe(result);
 return result;
}
\end{lstlisting}

\subsubsection{Dynamic Analysis}

Each extreme transformation $(m,m') \in U$ is observed and analyzed independently. After the instrumentation, all the test cases known to cover $m$ are executed. With this, we observe the original program state $\mathit{MS}_m$. The same is done with $m'$ to obtain $\mathit{MS}_{m'}$. If there is at least a property with a different value in both states, then the states are said to be different (\texttt{get\_diff} in \autoref{alg:diff}). The property points to the code location where the infection has manifested: one of the arguments, the receiver or the result of an invocation.

If no difference is observed, then the process has discovered a \textbf{no-infection} symptom for the given extreme transformation.

A method could be invoked several times by the same test case. In our experience, we have seen methods invoked thousands of times in a single test execution. During the dynamic analysis of this stage, each method invocation is uniquely identified by its order in the entire method invocation sequence. This identification differentiates the states from each invocation of $m$ and $m'$. The properties from one method invocation are distinguished from the properties of another method invocation. For example, the result value of the first invocation of $m$ is considered as a distinct property from the result value of the second invocation of $m$. With this, the state elements from the first invocation of $m$ are only compared to the same state elements form the first invocation of $m'$ and so on.

During the observations, there may be changes from one execution to the other, that are not due to the extreme transformations \cite{danglot_automatic_2019}. Such variations may be derived from random number generation, usage of system time, usage of temporary file paths, and they should be discarded. As a trivial example, in line \ref{line:example:time} of \autoref{list:example:time}, the value of \texttt{start} depends on the current system time. In order to identify such changing values, we execute the tests $N$ times when recording the states of the original methods and another $N$ times when recorded their transformed variants. $N$ is a parameter that can be controlled by the user with a default value of 10. 

\begin{lstlisting}[caption=Example of a test case using the system time. The value of \texttt{start} changes on every tests excution, captionpos=b, label=list:example:time]
@Test 
public void handlesManyChildren() {
 StringBuilder longBody = new StringBuilder(500000);
 for (int i = 0; i < 25000; i++) {
  longBody.append(i).append("<br>");
 }
 long start = currentTimeMillis(); %*\label{line:example:time}*)
 Document doc = Parser.parseBodyFragment(longBody);
 assertEquals(50000, doc.body().childNodeSize());
 assertTrue(currentTimeMillis() - start < 1000);
}
\end{lstlisting}

For the state comparison, we keep only the properties that had the same value across all $N$ executions. So, we actually consider as the state of the program only the set of state elements that remained unchanged across executions, as can be seen in the \texttt{get\_diff} function from \autoref{alg:diff}. 

If a difference is observed, we can collect the state property, that is, whether the infection can be observed in a field or the value of the result of the method, in one of the arguments or the receiver.

After this stage, the undetected transformations can be partitioned in two groups: those exhibiting a \textbf{no-infection} symptom, and those for which the test executions produce the infection of the immediate program state.

\subsection{Propagation detection}
\label{sec:propagation}

\begin{algorithm}
    \DontPrintSemicolon
    \KwData{$P$, $T$, \textit{infection}}
    \KwResult{\textit{no-propagation}, \textit{weak-oracle}}
  
    \SetKwFunction{Fpropagations}{find\_propagations}
    \SetKwFunction{Fobserve}{observe}
    \SetKwFunction{Fgetdiff}{get\_diff}
    \SetKwProg{Fun}{function}{does}{end}
  
    \Fun{\Fpropagations{$P$, $T$, \textit{infection}}}{
      \ForEach{ $ (m, m', \mathit{diff_m}) \in \mathit{infection} $ }
      {
        $T_m \gets$ \{ $t \in T$ : $t$ executes $m$ \} \;
        $T_{m,i} \gets$ instrument $T_m$ \;
        $\mathit{TS}_m \gets$ \Fobserve{$P$, $T_{m,i}$} \;
        $P' \gets$ replace $m$ by $m'$ in $P$ \;
        $\mathit{TS}_{m'} \gets$ \Fobserve{$P'$, $T_{m,i}$} \;
        $\mathit{diff_t} \gets$ \Fgetdiff{$\mathit{TS}_m$, $\mathit{TS}_{m'}$} \;
        \eIf{ $ \mathit{diff_t} \neq \emptyset $ } {
          \textit{weak-oracle} $\gets$ \textit{weak-oracle} $\cup \{ ((m, m'), \mathit{diff_t}) \} $ \;
        }{
          \textit{no-propagation} $\gets$ \textit{no-propagation} $\cup \{ ((m, m'), \mathit{diff_m}) \} $ \;
        }
      }
      \KwRet \textit{no-propagation}, \textit{weak-oracle}
    }
  
    \caption{Second stage of the process. This stage identifies \textbf{no-propagation} and \textbf{weak-oracle} symptoms}
    \label{alg:propagation}
\end{algorithm}

The goal of this second stage is to determine which program state infections, detected in the previous stage, are propagated to the test code. In this stage we observe the state of each test case covering these transformations. At the end, we are able to detect whether the infection could be observed in the existing test code (\textbf{weak-oracle}) or not (\textbf{no-propagation}). The main steps of this stage are outlined in \autoref{alg:propagation}.

\subsubsection{Instrumentation}

In this stage, the test states are also observed through source code instrumentation. Unlike in the previous stage, here we instrument the \java{} source code of the test cases. This instrumentation observes the test case state as defined in \autoref{def:test-state}. We instrument only the tests that cover the method of at least one extreme transformation for which a program infection was previously detected. The code is instrumented to observe the state of the values produced by all expressions and subexpressions in the code of the test case. This instrumentation amplifies the observation capabilities of the test cases, limited before only to the values being asserted by the developers.

 \autoref{list:instrumentation:propagation} shows the instrumentation applied to the \texttt{testAdd} test method from \autoref{list:example:test-suite}.
 
We exclude from the observation: constant expressions and expressions on the left side of an assignment. Generic test method and test classes, that is with type arguments, are not considered in our current implementation. The instrumentation  inserts \texttt{try...catch} blocks in the test code, in order to observe exceptions that could be thrown during test execution, as it is done in line \ref{line:example:exception}.

\begin{lstlisting}[caption=Test case instrumented to observe infection propagation, captionpos=b, label=list:instrumentation:propagation]
@Test
public void testAdd() throws Exception { %*\label{line:example:transformed:testAdd}*)
 try {
  VersionedSet list = 
   observe(new VersionedSet());
   observe(list).add(1); // State of list
   assertEquals(1, 
    observe( // Result of size
     observe(list).size() // State of list
    ));
 }
 catch(Exception exc) {
  observe(exc); %*\label{line:example:exception}*)
  throw exc;
 }
}
\end{lstlisting}

\subsubsection{Dynamic analysis}

The dynamic analysis in this stage is very similar to the infection detection. Each extreme transformation is analyzed in isolation. The test are executed $N$ times for the original code and $N$ times for the transformed code. Only the state elements that remained unchanged are kept for the state comparison.
Unlike the previous stage, here the states are recorded from the test code and the properties point to code locations where developers can actually insert new assertions.
If no difference was observed between the execution of the original method and the transformed method, then a \textbf{no-propagation} symptom is signaled. The previously detected infection did not propagate to an observable point in the test code. On the contrary, if a difference is observed, then we signal a \textbf{weak-oracle} symptom. The existing assertions do not notice the extreme transformation.

\subsection{Suggestion synthesis}
\label{sec:hints}

The third stage synthesizes the final suggestions. The suggestions are compiled into a human readable report. The report includes a detailed diagnosis of the symptom for each extreme transformation and an indication of the potential solution strategies. In this section we discuss how these reports are generated, the exact information they include and provide an example for each symptom.

\subsubsection{Suggestion for a no-infection symptom}

For a given undetected extreme transformation, if no immediate program infection is observed for all tests case, it means that the test input is not able to generate a state difference (\textbf{no-infection}). It also means that the method had no observable side-effects over the instance on which it was invoked or the arguments. The result value of the method is the same across all invocations. However, there are test cases that reach the method. 

Altering the existing input in these test cases could alter the program state so the method can return a different value or produce a different side effect.
So, the suggestion generated in the process tells developers \emph{to create new test cases from the ones covering the method} and alter their input to induce an observable effect. 

\begin{figure}
    \includegraphics[width=3.4in]{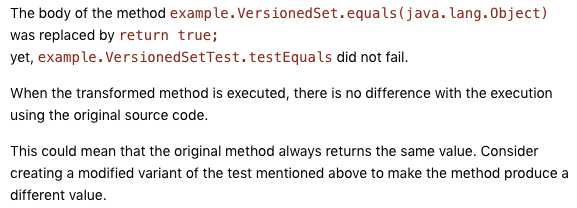}
    \caption{Suggestion synthesized for a \textbf{no-infection} symptom related to the \texttt{equals} method in our example.}
    \label{fig:sc-no-infection}
\end{figure}

Consider again the \texttt{equals} method in \autoref{list:example:class}. The suggestion given to the developer is to create a new test case to make the method return a different value using \texttt{testEquals} as a starting point. \autoref{fig:sc-no-infection} shows the report.

\subsubsection{Suggestion for a no-propagation symptom}

A \textbf{no-propagation} symptom unveils an immediate program infection that is not propagated to the test code. A different program state is observed after one or more invocations of the transformed method. However, no state difference is observed from the test cases. The effects of the program infection are masked at some point in the execution path. 
Observing the changed state requires a new invocation sequence able to propagate the infection to an observable point from the test. This means that \emph{new test cases are then required to detect these extreme transformations}. This the suggestion that we generate for these symptoms.

However, the method under study could be private and not directly accessible from the test cases. To address this problem, we find methods that are accessible from the test code, and that are as close as possible in an invocation sequence to the method we want to target.
For this we perform a static analysis in the code of the project that follows the invocation graph. The analysis produces the list of public or protected methods inside accessible classes that can be use to reach the method. 

The report for a \textbf{no-propagation} symptom includes a description of the transformation and the list of methods that should be targeted in the new test cases. If the method is already accessible then it is the only one listed as target. If it is not accessible, we include the list produced by the static analysis. The report also includes the test cases executing the method.

\begin{figure}
    \includegraphics[width=3.4in]{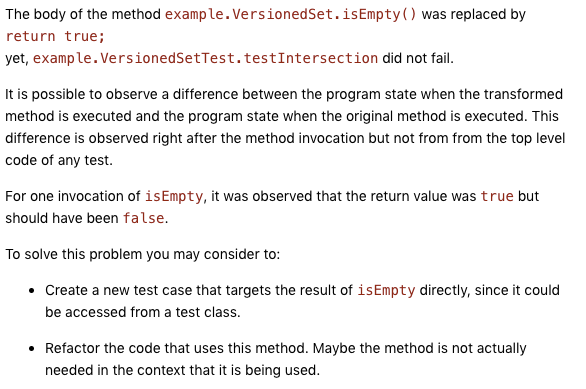}
    \caption{Suggestion synthesized for a \textbf{no-propagation} symptom related to the \texttt{isEmpty} method in our example.}
    \label{fig:sc-no-propagation}
\end{figure}

In our example, the \texttt{isEmpty} method has a \textbf{no-propagation} symptom. As, it is accessible from the test code, the suggestion is to create a new test case targeting \texttt{isEmpty} directly.
\autoref{fig:sc-no-propagation} shows the report.

\subsubsection{Suggestion for a weak-oracle symptom}

If an infection propagation is detected (\textbf{weak-oracle}), it is visible in the result of an expression in the test code. In the previous stage we have recorded the exact code location of this expression.

The \textbf{weak-oracle} suggestion is \emph{to add an assertion at the right location in the test code}. We even provide the developer with the value to be asserted. 

The state difference could be the result of the expression itself if it is a primitive type or a string. But, if the expression returns an object, the difference may be observed in one of its fields. For instance, in our example from \autoref{list:example:test-suite} in line \ref{line:example:observed-propagation} the result of the expression at the right of the assignment is a \texttt{VersionedSet}. With the original code, the value of the \texttt{version} field should be \texttt{1} while, with the transformed method it is observed to be \texttt{0}.

If the state difference is observed through an accessible field of the result of the expression or the result itself, the suggestion is to create an assertion targeting this value.

If the field is not accessible, (\ie{} it is private) further actions are required. For this, we perform a static analysis to find methods that use the identified field. These methods could be not accessible from the test code. So, we find accessible class members invoking these methods, in the same way it is done for the \textbf{no-propagation} reports. The final suggestion is to assert the result and side effects of the final set of methods, invoked in the result of the initially identified expression.

\begin{figure}
    \includegraphics[width=3.4in]{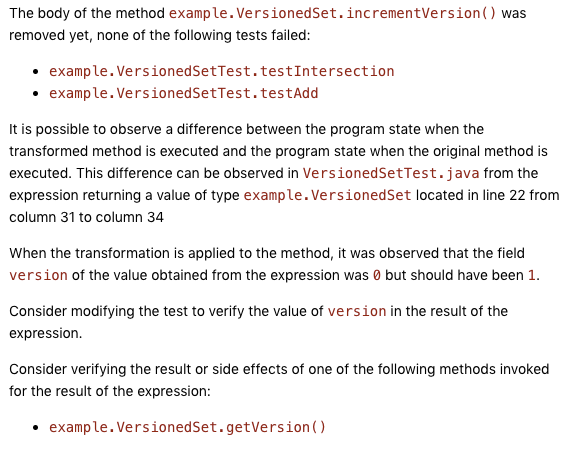}
    \caption{Suggestion synthesized for a \textbf{weak-oracle} symptom related to the \texttt{getVersion} method in our example.}
    \label{fig:sc-weak-oracle}
\end{figure}

In \autoref{list:example:class}, the process suggests the addition of a new assertion targeting the \texttt{getVersion} method invoked in the list instance of line \ref{line:example:insert-assertion}. 
\autoref{fig:sc-weak-oracle} shows the synthesized suggestion.

\subsection{Implementation}

We have implemented our proposal in an open-source tool which we named \tool. The tool has been conceived as a \maven plugin and it is able to target \java programs that use \maven as main build system. The code, and all data related to the current work are available from \github{} \footnote{\url{https://github.com/STAMP-project/reneri}} \footnote{\url{https://github.com/STAMP-project/descartes-amplification-experiments}}. \tool relies on Javassist 3.24.1 \cite{chiba_load-time_2000} for bytecode instrumentation, Spoon 7.1.0 \cite{pawlak_spoon_2016} for static code analysis and \java source code transformation and \descartes 1.2.5 \cite{vera_descartes_2018} to apply extreme transformations. All stages in the process described before are exposed as \maven goals. Apart from human readable reports, \tool also generate files more suitable for automatic analysis that could be used by external tools.

\section{Experimental Evaluation}

We assess our proposal based on a set of research questions. In this section we present these questions, the projects we used as study subjects and the results we obtained.

\subsection{Research questions}

\newcommand{\rqinfecttext}{\textit{%
To what extent does the execution of an extreme transformation infect the immediate program state?}}
\newcommand{\rqpropagatetext}{\textit{%
To what extent can test cases propagate the effects of extreme transformations to the top level test code?}}
\newcommand{\rqevaltext}{\textit{%
Are the suggestions synthesized by \tool valuable for the developers?}}
\newcommand{\rqsolvetext}{\textit{Can developers leverage test improvement tools to deal with undetected extreme transformations?}}

\begin{enumerate}[label=\textbf{RQ\arabic*}, wide, labelwidth=!, labelindent=0pt]

	\item \label{rqinfect} \rqinfecttext{}
	
	With this question we quantify 1) how frequent the \textbf{no-infection} symptom is and 2) how often an infection can be observed. We collect the number of extreme transformations exhibiting an infection across projects.

    \item \label{rqpropagate} \rqpropagatetext{}
	
	With this question we quantify how frequently \textbf{no-propagation} and \textbf{weak-oracle} symptoms appear across projects.  \textbf{weak-oracle} symptoms might represent assertions missed by developers. \textbf{no-propagation} symptoms may require the creation of new test cases.
	
    \item \label{rqeval} \rqevaltext{}

	In this question we empirically assess the suggestions generated by \tool. The goal is to know if explaining the undetected extreme transformations can actually help developers to improve their test cases. 

	\item \label{rqsolve} \rqsolvetext{}
	
	With this question we explore if developers can leverage state-of-the-art automatic test generation and test improvement tools. We also quantify the effectiveness of these tools against \textbf{no-infection}, \textbf{no-propagation} and \textbf{weak-oracle} symptoms.
	
\end{enumerate}

\subsection{Study subjects}

We select a set of 15 open-source projects as study subjects to answer our research questions. All projects use \maven{} as the build system, are written in \java{} (8 or lower) and use \junit{}(4.12 or lower). We selected these projects systematically as follows. We started from 51 single-module projects, matching the above conditions and studied by the authors of \cite{urli_how_2018} and \ourpreviouspaper{}. 
We focus on single-module projects as they have a very well defined structure and all unit test cases are well located. This eliminates the need to deal with the idiosyncrasies from specific projects.

For these 51 initial projects, we performed the following actions:
\begin{enumerate*}
	\label{workflow:discovery}
	\item Clone the repository
	\item Switch to the commit of the last stable release (according to the \github{} API) or the latest commit if the API produced no result
	\item Build the project using \texttt{mvn clean test}
	\item Execute \pit{} using the \descartes{} extension to find undetected extreme transformations.
\end{enumerate*}

The last step  computes the set of undetected extreme transformations and the test cases reaching these transformations. The full workflow was successful in only 25 projects. For the other 26 projects the main reasons of failure were compilation errors, failing tests or specific set-up requirements or dependencies external to the \maven{} ecosystem.

From these 25 projects, we kept those having at least one undetected extreme transformation. We discarded projects with more than 100 application methods with at least one undetected extreme transformation. With this we eliminated 6 projects where all extreme transformations were detected and 2 that had more than 100 affected methods. 2 additional projects were discarded due to technical incompatibilities with our implementation: one of them widely uses generic classes in the test code,  which we do not target in our current implementation. The testing configuration of the other project clashed with the way we log the program state information during test execution.

\autoref{tab:projects} shows the list of projects used in the present study. The columns in the table show the number of lines of the main source code (LOCs) as measured by \textit{cloc}\footnote{\url{https://github.com/AlDanial/cloc}}, the number of lines for the test code, the number of test cases in the project as reported by \maven{} with the execution of \texttt{mvn test}, the number of application methods for which at least one extreme transformation was not detected and the number of undetected extreme transformations for the entire project. 

The last three columns provide statistics about how many test cases execute each method included in the \emph{Methods} column:. the minimum, maximum and average number of tests cases that execute a method.

\begin{table*}
	\caption{Projects used as study subjects. The first column is the name of the project. \emph{App LOC} and \emph{Test LOC} show the lines of code for the application code and the test code respectively. \emph{Tests} Shows the number of test cases as reported by Maven. \emph{Methods} shows the number of application methods with at least one undetected extreme transformation. \emph{Transformations} shows the number of undetected extreme transformations in the project. \emph{Min \#Test}, \emph{Max \#Test} and \emph{Avg \#Test} refer to the minimum, maximum and average number of test cases in \emph{Test} that execute each method in \emph{Method}}
	\label{tab:projects}
	\begin{adjustbox}{max width=\textwidth,center=\textwidth}
	\begin{tabular}{lrrrrr|rrr}
		\toprule  
		Project                                     & {App LOC} & {Test LOC} & {Tests} & {Methods} & {Transformations} & Min \#Test & Max \#Test & Avg  \#Test \\ 
		\midrule
		\texttt{jpush-api-java-client}              &      3667 &       3112 &      72 &         2 &                 2 &          1 &          5 &          3 \\ 
		\texttt{commons-cli}                        &      2800 &       4287 &     471 &         3 &                 5 &          8 &        154 &         61 \\ 
		\texttt{jopt-simple}                        &      2410 &       6940 &     838 &         3 &                 7 &          1 &         77 &         26 \\ 
		\texttt{yahoofinance-api}                   &      2925 &        453 &      15 &         3 &                 5 &          3 &          4 &          4 \\ 
		\texttt{gson-fire}                          &      1665 &       1644 &      79 &         4 &                 4 &          2 &         17 &          7 \\ 
		\texttt{j2html}                             &      3460 &       1250 &      51 &         5 &                 5 &          1 &          1 &          1 \\ 
		\texttt{spring-petclinic}                   &       731 &        687 &      40 &         6 &                 6 &          2 &         11 &          5 \\ 
		\texttt{javapoet}                           &      3363 &       5334 &     340 &        10 &                10 &          1 &        104 &         12 \\ 
		\texttt{eaxy}                               &      3599 &       1670 &     204 &        11 &                14 &          1 &         63 &         18 \\ 
		\texttt{java-html-sanitizer}                &      7662 &       5944 &     260 &        13 &                14 &          1 &        134 &         44 \\ 
		\texttt{cron-utils}                         &      4544 &       5202 &     466 &        16 &                16 &          1 &        207 &         49 \\ 
		\texttt{commons-codec}                      &      8241 &      11957 &     889 &        20 &                22 &          1 &         96 &         21 \\ 
		\texttt{jsoup}                              &     12015 &       7911 &     680 &        33 &                38 &          1 &        483 &         50 \\ 
		\texttt{TridentSDK}                         &      5544 &       1670 &     127 &        35 &                46 &          1 &         17 &          3 \\ 
		\texttt{jcodemodel}                         &     13752 &       1544 &      68 &        95 &               118 &          1 &         50 &         11 \\ 
		\midrule
		Global                                      &     76378 &      59605 &    4600 &       259 &               312 &          1 &        483 &         20 \\
		\bottomrule
	\end{tabular}
\end{adjustbox}
\end{table*}

All projects included in the study are of small to medium size (731 to 13752 LOCs). In seven cases, the size of the test suite is comparable or larger to the size of the application code. 
The size is not always correlated to the number of test cases which range from 15 to 889. There are 260 methods methods with at least one undetected extreme transformation, and they are distributed as follows: from 2 to 95 per project. The total number of transformations studied is \tuet\ ranging from 2 to 118 per project. In most projects there is at least one transformation that is covered by only one test case. In four projects all the transformations  are executed by more than one test case and in \texttt{commons-cli} the methods are covered by at least eight test cases. Notably, in five projects, one of the identified methods is covered by more than 100 test case: in \texttt{jsoup}, one transformation is covered by 483 test cases. The transformations are covered on average by 20 test cases.

\subsection{\ref{rqinfect}: \rqinfecttext{}}
\label{rqinfect:answer}

To answer \ref{rqinfect} we executed the first stage of \tool and collected the number of \textbf{no-infection} symptoms for all our 15 study subjects. 

We observe an immediate program infection for \tpi\ out of the \tuet\ undetected extreme transformations. That is, in \percentage\tpi\tuet\ of the cases, the existing test inputs do trigger an infection of the local program state at the transformation point. The rest \tni\ (\percentage\tni\tuet) does not provoke a program infection. So, the execution of the majority of transformations in fact provoke a program infection.

\begin{figure}
    \centering
    \includegraphics[width=2.5in]{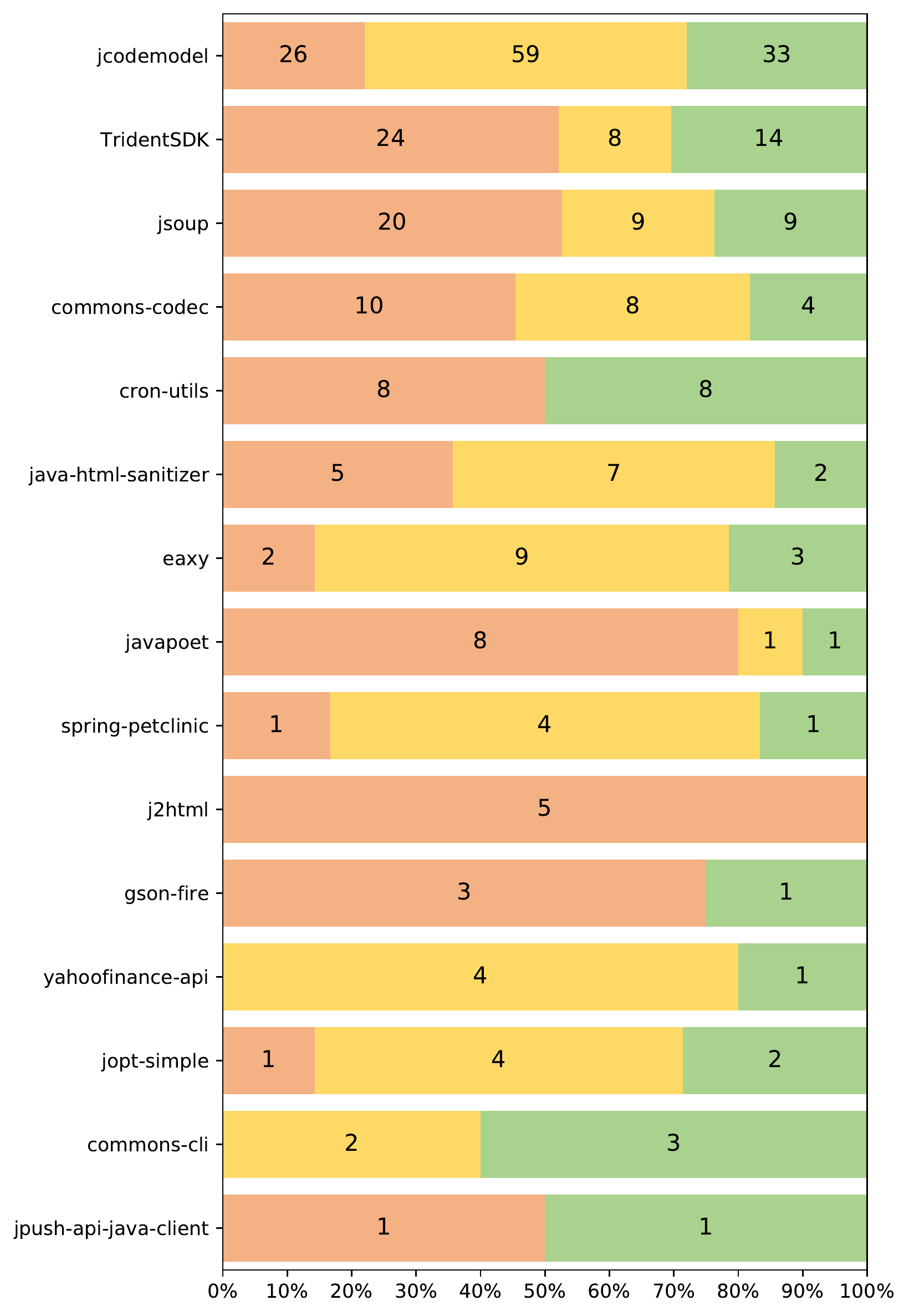}
    \caption{Number and proportion of \textbf{no-infection} in orange, \textbf{no-propagation} in yellow and \textbf{weak-oracle} in green found for each project.}
    \label{fig:rip-plot}
\end{figure}

\autoref{fig:rip-plot} shows the number and proportion of the three symptoms: \textbf{no-infection}, \textbf{no-propagation} and \textbf{weak-oracle} found for each study subject. 
For instance, for project \texttt{jpush-api-java-client} there is 50\% of cases with \textbf{no-infection} and 50\% of cases with \textbf{weak-oracle}.
In all but three projects the \textbf{no-infection} symptoms are less than 52\% of all symptoms discovered. In two projects, \texttt{commons-cli} and \texttt{yahoofinance-api} the execution of all extreme transformations infected the program state. On the contrary, in \texttt{j2html} all the symptoms are \texttt{no-infection}. So, we observe that the proportion of symptoms is different from one project to another.

A \textbf{no-infection} symptom occurs when the effects and result of the transformed method do not differ from the effects and results of the original method, with respect to the existing testing input. As an example, one of the methods in \texttt{j2html} is an \texttt{equals} method executed by only one test case where it returns \texttt{true}. There is no test case where the method should return false, that is, it is never presented with objects that are not equal. 

The other four methods in the same project are all void and are entitled to perform a sanitization in the fields of the class instance, if needed. These four methods are executed by only one test case whose input does not contain a string requiring the sanitization. Thus all these methods are executed and the expected program state is the same under the extreme transformations. So, here, even when the test code executes these methods, the input does not trigger a behavior where their effects can be manifested. 

One common cause that prevents an immediate program infection is related to boolean methods. In many cases these methods are tested to return only one value: always true or always false. In such a situation, the transformed method will have the same result in all invocations as the original method. This produces a \textbf{no-infection} symptom. In fact, in our study, \tbool\ undetected extreme transformations were discovered in boolean methods. \tboolni\ of them do not provoke a program infection. Notably, 19 were created in \texttt{equals} methods and none of them induced a program infection. 

So, boolean methods are related to a large portion of undetected extreme transformations (\percentage\tbool\tuet) and half of them exhibit a \textbf{no-infection} symptom.

Another common scenario comes from methods checking preconditions that are not tested with corner cases. \autoref{list:rqinfect:eaxy} shows one of those methods. As can be seen, these are void methods that check the value of a boolean expression (line \ref{line:precondition}), which is the precondition. If the expression is false, then an exception is thrown as in line \ref{line:precondition:exc}. So, apart from the exception, these methods have no other effect. It is common that developers do not create test cases with inputs that do not meet the precondition.

\begin{lstlisting}[caption=A \textbf{no-infection} example. The precondition is always satisfied in the test cases. the corner case is not verified, captionpos=b, label=list:rqinfect:eaxy]
private void checkMaxOneMatch() {
 if (size() <= 1) return; %*\label{line:precondition}*)
 String message = ...
 throw new IllegalArgumentException(message); %*\label{line:precondition:exc}*)
}
\end{lstlisting}

\subsubsection*{Where are the infections detected?}

\tool observes program infections in the result of the invocation of the transformed method, its arguments after the invocation and the instance on which the method was invoked. In this section we explore how frequently the infection is observed at each location.

The result value can only be observed in non-void methods. In this study we analyze \tuet\ extreme transformations: 254 of them were created in non-void methods. 176/254 transformations in non-void methods are related to an infection and 167/176 (95\%) of these infections are observed in the state of the result value. So, in non-void methods, the infection is observed in the result value in a vast majority of cases.

An infection in the arguments can only be observed for parameterized methods. Our study includes 227 transformations in methods with parameters. 144/227 of them caused an infection. We observe that 37/144 (26\%) of these infections are detected in the state of the arguments. That corresponds to the best practice that the returned value stores the effect, not the argument.

An effect in the receiver instance can only be observed in non-static methods: 251 transformations in our study were created in instance methods. 160/251 of them provoked an infection and 69/160 (43\%) can be detected in the state of the receiver instance.

In global terms, of the \tpi\ transformations that caused an infection, 84\% (167) can be observed in the state of the result value of the method, 19\% (37) can be observed in the state of the arguments and 35\% (69) can be detected in the state of the receiver instance. While, these numbers are more a reflection of the coding practices than of extreme transformations and even testing, they still provide valuable insights for developers who wish to understand undetected transformations.


\autoref{tab:infection-places} shows the details for the study subjects. The columns are divided in three groups. The first group corresponds to the transformations created in non-void methods, how many of them caused an infection and how many of them were detected in the state of the result value. The two other groups show analogous numbers for transformations in methods with parameters, non-static methods and whether the transformations can be detected in the state of the arguments and the receiver instance. The numbers for each project does not differ much from the general view. Only \texttt{spring-petclinic} shows more infections in the state of the arguments (4) less in the receiver instance (2) and none in the state of the result value.

\begin{table*}
	\caption{Number of times extreme transformations were detected in the state of the result value, the arguments or the receiver instance. The table is divided in three groups: non-void methods, parameterized methods and non-static methods. Each group shows the total number of transformations, how many of them provoke an infection and if they were detected in the result, the arguments or the receiver respectively.}
	\label{tab:infection-places}
	\begin{adjustbox}{max width=\textwidth,center=\textwidth}
	\begin{tabular}{l|rrr|rrr|rrr}
		\toprule  
    & 
    \multicolumn{3}{c}{Transformations in non-void methods} &
    \multicolumn{3}{c}{Transformations in parameterized methods} &
    \multicolumn{3}{c}{Transformations in non-static methods} \\
		Project & 
		Total & 
		Infection &
		In result value &
		Total &
		Infection &
		In parameters &
		Total &
		Infection &
		In receiver instance \\
		\midrule

        jpush-api-java-client &   2 &   1 &   1 &   1 &   0 &  0 &   1 &   0 &  0 \\
        commons-cli           &   5 &   5 &   5 &   4 &   4 &  0 &   2 &   2 &  0 \\
        jopt-simple           &   7 &   6 &   6 &   7 &   6 &  0 &   6 &   6 &  0 \\
        yahoofinance-api      &   5 &   5 &   5 &   5 &   5 &  0 &   0 &   0 &  0 \\
        gson-fire             &   4 &   1 &   1 &   4 &   1 &  1 &   3 &   1 &  1 \\
        j2html                &   1 &   0 &   0 &   1 &   0 &  0 &   5 &   0 &  0 \\
        spring-petclinic      &   2 &   1 &   0 &   6 &   5 &  4 &   6 &   5 &  2 \\
        javapoet              &   9 &   2 &   2 &   9 &   1 &  1 &   8 &   1 &  1 \\
        eaxy                  &  12 &  11 &  10 &   7 &   5 &  0 &  14 &  12 &  5 \\
        java-html-sanitizer   &   9 &   7 &   5 &  10 &   5 &  2 &   7 &   6 &  2 \\
        cron-utils            &  14 &   8 &   8 &  16 &   8 &  0 &  14 &   7 &  2 \\
        commons-codec         &  22 &  12 &  12 &  20 &  10 &  2 &  17 &   8 &  2 \\
        jsoup                 &  27 &  15 &  15 &  24 &  13 &  3 &  33 &  16 &  6 \\
        TridentSDK            &  44 &  22 &  21 &  33 &  17 &  2 &  40 &  20 &  6 \\
        jcodemodel            &  91 &  80 &  76 &  80 &  64 & 22 &  95 &  76 & 42 \\

		\midrule
        Total                 & 254	& 176 &	167 & 227 & 144	& 37 & 251 & 160 & 69 \\
		\bottomrule
 
	\end{tabular}
	\end{adjustbox}
\end{table*}

\takeaway{rqinfect}{The existing tests infect the immediate program state at the transformation point in \percentage\tpi\tuet\ of the cases. This indicates that undetected extreme transformations are mostly due to a lack of observation in the test cases. 84\% of transformations infecting the program state result in a different return value for the transformed method.}

\subsection{\ref{rqpropagate}: \rqpropagatetext{}}
\label{rqpropagate:answer}

To answer this question we execute the second stage of our process and collect all \textbf{no-propagation} and \textbf{weak-oracle} symptoms. We observe an infection that successfully propagates (i.e., a \textbf{weak-oracle})  for \twor\ extreme transformations, which represents 42\% of the 198 transformations causing a program infection (and 27\% of all transformations in the study). The other 115 transformations that can infect the program are diagnosed as \textbf{no-propagation}, which represents 58\% of the transformations that do infect the program state (37\% of all transformations). A general observation is that a weak oracle is clearly not the main symptom that prevents the test suite from detecting the extreme transformations.

Here again we notice differences among projects. In \autoref{fig:rip-plot} we notice four projects that do not have a single case of \textbf{no-propagation}. In three of them, all program infections reached the test code at locations that can be verified. The other, \texttt{j2html} had no program infection, which was discussed previously. In 11 projects, the \textbf{weak-oracle} symptoms are close or below 30\% of all symptoms detected. In \texttt{commons-cli}, \texttt{jpush-api-java-client} and \texttt{cron-utils} these symptoms were majority or reached the 50\% of all symptoms discovered for the project.

We hypothesize that the distance to the test code might be a factor that influences \texttt{no-propagation} symptoms. If the methods are far from the test code in the invocation sequence, then their effects are more likely to get hidden or lost before reaching the test code. To check this hypothesis, we compute the stack distance from the method to the test case for all transformations. We consider the stack distance to be the number of activation records in the method call stack between the activation record of the test case and the activation record of the method related to the transformation. If the method is invoked more than once in the test suite, we keep the lowest stack distance.

\begin{figure}
    \centering
    \includegraphics[width=2.5in]{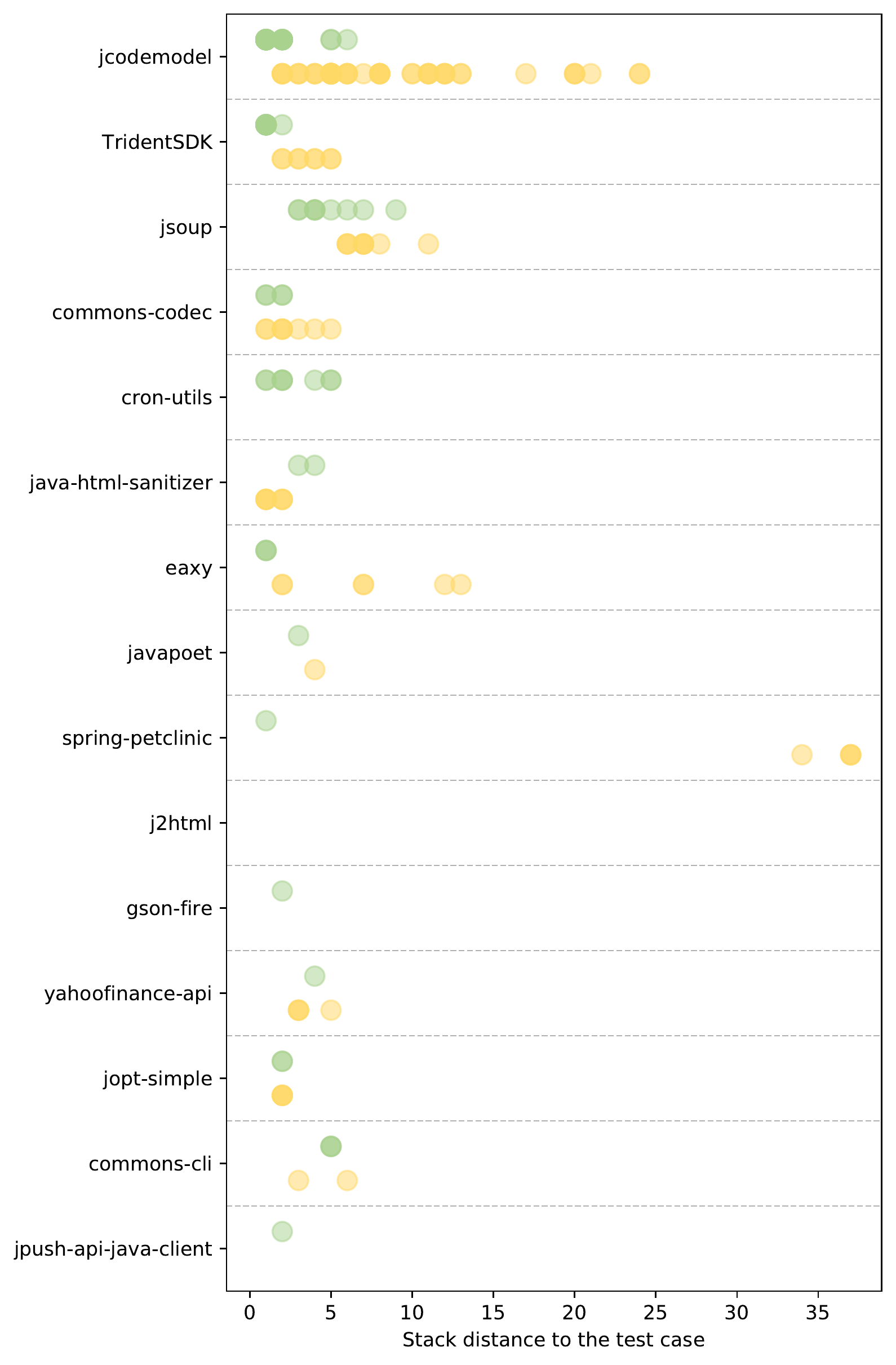}
    \caption{Stack distance from the method to the test case for \textbf{no-propagation} symptoms (yellow) and \textbf{weak-oracle} symptoms (green).}
    \label{fig:distances}
\end{figure}

\autoref{fig:distances} plots these distance measures between the test and the transformed method. Each point in the graph corresponds to an extreme transformation in the study. Transformations with \textbf{no-propagation} symptoms are shown in  yellow and transformations with \textbf{weak-oracle} symptoms are shown in green. For example, the green point in  \texttt{commons-cli} corresponds to an extreme transformation with a \textbf{weak-oracle} symptom, whose method is five activation records away from an activation record of a method in a test class, when the tests are executed. Darker points in the graph result from the agglomeration of several extreme transformations with the same stack distance.
We observe that, in most cases, the stack distance of \textbf{weak-oracles} tends to be lower than the  distance for \textbf{no-propagation} symptoms. This is evidence that confirms our hypothesis: transformations that occur far from the tests tend to propagate less than the others. Smaller projects tend to behave differently. The few \textbf{weak-oracle} cases that have higher stack distances occur on small projects, e.g., \texttt{java-html-sanitizer} or \texttt{jopt-simple} where they have the same distance. Larger projects,\texttt{jsoup}, \texttt{TridentSDK}, \texttt{jcodemodel} show a more significant tendency to the stack distance difference between these two symptoms.

\subsubsection{\textbf{no-propagation} examples}

An infection is not propagated when the code executed between the method invocation and the top level code of the test case \textit{masks} the state infection. In many cases this happens for methods that are reached through a long sequence of invocations. An example from \texttt{commons-cli} is given in \autoref{list:rqobserve:commons-cli}: when the body of \texttt{isLongOption} (line \ref{line:rqobserve:isLongOption}) is replaced by \texttt{return false}, the test suite does not fail. This method is reached through a sequence of five invocations starting from \texttt{parse} (line \ref{line:rqobserve:parse}). This sequence of invocations is triggered only when specific conditions are met (line \ref{line:rqobserve:condition}). \texttt{isLongOption} is invoked 32 times by seven test cases in the test suite. The method is expected to return true in only one invocation. When the method is transformed to return only false, the infection is observed for the invocation in which it should have returned true. However, in this case, the \texttt{isShortOption} (line \ref{line:rqobserve:option-condition}) method is executed and returns true. Therefore, the return value of \texttt{isOption} (line \ref{line:rqobserve:isOption}) does not change in comparison with the expected result. Since these methods have no other effects, the infection is not propagated.

\begin{lstlisting}[caption=An example of \textbf{no-propagation} in \texttt{commons-cli} where the execution masks the effects of the transformed method, captionpos=b, label=list:rqobserve:commons-cli]
public class DefaultParser {
 private boolean isLongOption(String token)  {...} %*\label{line:rqobserve:isLongOption}*) 

 private boolean isOption(String token){ %*\label{line:rqobserve:isOption}*)
  return isLongOption(token) || isShortOption(token); } %*\label{line:rqobserve:option-condition}*)

 private boolean isArgument(final String token) { %*\label{line:rqobserve:isArgument}*)
  return !isOption(token) || isNegativeNumber(token); }

 private void handleToken(final String token) {
 ...
  else if (currentOption != null && currentOption.acceptsArg() && isArgument(token)) %*\label{line:rqobserve:condition}*)
  ...
 }

 public CommandLine parse(...){ %*\label{line:rqobserve:parse}*)
  ...
  if (arguments != null)
   for (final String argument : arguments) 
    handleToken(argument);
  ...
 }
}
\end{lstlisting}

A \textbf{no-propagation} occurrence might signal the need for code refactoring. In \texttt{commons-cli} we found that the body of \texttt{hasValueSeparator} declared in the \texttt{Option} class (line \ref{line:rqobserve:hasValueSeparator} in \autoref{list:rqobserve:commons-cli-2}) could be replaced by \texttt{return true} and, for the test cases in which the method should have returned false, the change in the program state  does not reach the top level test code. It is interesting to notice that this is a public method, not directly assessed by any test case and it is used only by a private method in the same class. In the code of \texttt{processValue}, if the instance of \texttt{Option} has a value separator, some specific actions are taken. These actions depend on the presence of such a separator. If \texttt{hasValueSeparator} returns the wrong value, the separator is not found in the given parameter, and the actions are never taken (see the \texttt{while} loop in line \ref{line:rqobserve:while}). So, in this example, \texttt{hasValueSeparator} has no effect whatsoever and the code could be refactored. Another way to solve the problem is to directly assess the method, as it is public.

\begin{lstlisting}[caption=An example of a \textbf{no-propagation} symptom in \texttt{commons-cli} that might require refactoring, captionpos=b, label=list:rqobserve:commons-cli-2]
public boolean hasValueSeparator() {%*\label{line:rqobserve:hasValueSeparator}*)
  return valuesep > 0;
}

private void processValue(String value) %*\label{line:rqobserve:processValue}*)
{
 if (hasValueSeparator()) {
  char sep = getValueSeparator();
  int index = value.indexOf(sep);
  while (index != -1) { %*\label{line:rqobserve:while}*)
   if (values.size() == numberOfArgs - 1) break;
   add(value.substring(0, index));
   value = value.substring(index + 1);
   index = value.indexOf(sep);
  }
 }
 add(value);
}
\end{lstlisting}

In general, a \textbf{no-propagation} symptom might be solved with the creation of new test cases which are closer, in the invocation chain, to the method in question. This increases the chances of immediate program infections to propagate to the top level code of the test case.

\subsubsection{\textbf{weak-oracle} examples}

When the infection is propagated to the test code we discover a \textbf{weak-oracle} symptom. The infection propagations are observed in the result value of the expression in the code of the test cases.

\autoref{list:weak-public} exposes an example of a \textbf{weak-oracle} symptom in \texttt{jcodemodel}. 

The \texttt{annotate} methods of the class \texttt{JPackage}, declared in lines \ref{line:weak-public:second-method} and \ref{line:weak-public:first-method} have both a such a symptom. The body of these two methods can be replaced by \texttt{return null} and the test suite executes without noticing the change. These methods are executed by the \texttt{testPackageAnnotation} test case shown in line \ref{line:weak-public:test}. This test seems to be designed to actually verify the package annotation functionality. Both transformations do infect the program state, as they make the method return null, which is not the expected value. The infection reaches the test code in line \ref{line:weak-public:annotate-invocation}, where the method is invoked and its result could be directly asserted. The transformation to the method in line \ref{line:weak-public:first-method} can be also observed in line \ref{line:weak-public:model-observation}. This method has a side effect on the \texttt{JCodeModel} instance by adding a reference to the annotation class. In the \texttt{JCodeModel} class, these references are stored in the \texttt{m\_aRefClasses} field (line \ref{line:weak-public:references}). Our process notices that this field does not have the expected size when the transformation is executed, as it misses a reference to \texttt{Inherited}. The oracle in the same line \ref{line:weak-public:model-observation} checks only if the code model is syntactically correct. It does not check if the package has been annotated.

\begin{lstlisting}[caption=Example of \textbf{weak-oracle} symptoms in public methods in \texttt{jcodemodel}, captionpos=b, label=list:weak-public]
public class JPackage ... { 
 private final JCodeModel m_aOwner; %*\label{line:weak-public:model-declaration}*)
 private List<JAnnotationUse> m_aAnnotations;

 public JAnnotationUse annotate (AbstractJClass aClazz) {%*\label{line:weak-public:second-method}*)
  JCValueEnforcer.isFalse(isUnnamed (), "...");
  if (m_aAnnotations == null)
   m_aAnnotations = new ArrayList<>();
  JAnnotationUse a = new JAnnotationUse(aClazz);
  m_aAnnotations.add(a);
  return a;
 }

 public JAnnotationUse annotate (Class aClazz) { %*\label{line:weak-public:first-method}*)
  return annotate(m_aOwner.ref(aClazz));
 }
 ...
}

public class JCodeModel {
 private Map<Class<?>, JReferencedClass> m_aRefClasses = new HashMap<> (); %*\label{line:weak-public:references}*)
 ...
}

@Test 
public void testPackageAnnotation () { %*\label{line:weak-public:test}*)
 JCodeModel cm = new JCodeModel();
 cm._package("foo").annotate(Inherited.class); %*\label{line:weak-public:annotate-invocation}*)
 CodeModelTestsHelper.parseCodeModel(cm); %*\label{line:weak-public:model-observation}*)
}
\end{lstlisting}

Testability issues may prevent \textbf{weak-oracle} symptoms to be solved without modifying or refactoring the main code base. \autoref{list:rqpropagate:gson-fire} shows the example of a private void method named \texttt{getFromCache} in \texttt{gson-fire}. It is used by a public method (lines \ref{line:rqobserve:cache-usage-1} and \ref{line:rqobserve:cache-usage-2}) to retrieve a cached result. Under extreme transformations the instructions of this method are replaced by \texttt{return null}. The immediate program state is infected for some invocations of the method where the result value should be non-null.  When the cached result is null, the actual value is computed once again, therefore the returned value of the public method is the same. The effects are propagated and observed in the \texttt{cache} field (line \ref{line:rqobserve:cache-usage-3}). Under the transformation no value is ever cached and the collection remains empty when it should contain some elements. However, there is no way to verify the effects of this method from the test code. The process is able to get the size of \texttt{cache} via reflection, but the field is not accessible. In order to catch the transformation, it is required to verify the content of \texttt{cache} with an extra accessor or changing its visibility.

The effects could be observed in the \texttt{cache} field . However, this field is private and cannot be observed from any test case. Moreover, it can not be solved with the creation of any new test case. It requires a code modification.

We inspected the 58 transformations of the first nine projects in \autoref{tab:projects}. For each transformation we manually wrote a test case trying to detect testability issues. We found only two cases, one of them is the method exposed in \autoref{list:rqpropagate:gson-fire}. So, even when we observed concrete cases, lack of testability does not appear to be a major influence for undetected extreme transformations.
    
\begin{lstlisting}[caption=Example of a testability issue preventing the solution of a \textbf{weak-oracle} symptom in \texttt{gson-fire}, captionpos=b, label=list:rqpropagate:gson-fire]
public abstract class AnnotationInspector {
 private ConcurrentMap cache = new ConcurrentHashMap();
 
 private Collection getFromCache(Class clazz, Class annotation) { %*\label{line:rqobserve:getFromCache}*)
  Map annotationMap = cache.get(clazz);
  if(annotationMap != null){
   Collection methods = annotationMap.get(annotation);
   if(methods != null)
    return methods;
  }
  return null;
 }

 public Collection getAnnotatedMembers(Class clazz, Class annotation){
  if(clazz != null) {
   Collection members = getFromCache(clazz, annotation); %*\label{line:rqobserve:cache-usage-1}*)
   if (members != null)
    return members;
   //Cache miss
   members = getFromCache(clazz, annotation); %*\label{line:rqobserve:cache-usage-2}*)
   if (members == null) {
    ...
    ConcurrentMap storedAnnotationMap = cache.putIfAbsent(clazz, newAnnotationMap); %*\label{line:rqobserve:cache-usage-3}*)
    ...
   }
  }
  return Collections.emptyList();
 }
}
\end{lstlisting}

\takeaway{rqpropagate}{ 42\% of the transformations infecting the program state, propagate to an observable point. In general, methods with \textbf{weak-oracle} symptoms are closer to the test cases than \textbf{no-propagation} in the invocation sequence. Transformations that do not propagate may require new test cases closer to the method. Testability issues may prevent \textbf{weak-oracle} symptoms to be solved without refactoring. }

\subsection{\ref{rqeval}: \rqevaltext{}}
\label{rqeval:answer}

With \ref{rqeval} we perform a qualitative evaluation of the suggestions synthesized by \tool. We want to know if developers find these suggestions relevant and helpful to fix undetected transformations.
To answer this question we target four open-source projects for which we are able to directly consult developers who have a strong knowledge about the code.

For each project we select up to six undetected extreme transformations for which \tool generated a suggestion. We manually select these transformations, so that they represent interesting testing cases for the developers. Table \autoref{tab:projects:eval-subjects} shows the projects and the symptoms of the selected issues.

For each issue we create a form containing 
\begin{enumerate*}[label=(\roman*)]
    \item a link to the automatically generated report (similar to those shown in Figures \ref{fig:sc-no-infection}, \ref{fig:sc-no-propagation} and \ref{fig:sc-weak-oracle}); and 
    \item a set of questions \footnote{The questionnaire can be consulted in \url{https://github.com/STAMP-project/descartes-amplification-experiments/blob/master/validation/form.md}}.
\end{enumerate*}

\begin{table}
	\caption{Projects involved in the empirical validation with developers}
	\label{tab:projects:eval-subjects}
	\centering
	\begin{tabular}{lrrrrr}
		\toprule  
		Project & 
		Issues & 
		\textbf{ni} &
		\textbf{np} & 
		\textbf{wo} & 
		\\
		
		\midrule
		\texttt{Funcon4J}             &  5 &  2 &   3 &  0 \\
		\texttt{greycat}              &  6 &  2 &   2 &  2 \\
		\texttt{sat4j-core}           &  6 &  2 &   2 &  2 \\
		\texttt{xwiki-commons-job}    &  6 &  2 &   3 &  1 \\
		\midrule
		Total                         & 23 &  8 &  10 &  5 \\
		\bottomrule
	\end{tabular}

\end{table}

\subsubsection{Qualitative feedback from the developers}

\autoref{tab:feedback} summarizes the developers'  feedback. We divide the answers from the developers according to the three types of symptom. Of the 23 issues that were discussed, 18 were considered as relevant or of medium relevance. The developers considered the report to contain helpful information in 18 cases, and even thought that the exact testing solution was given in 4 cases. In all cases, the developers, could emit a verdict about all the testing issues in around 30 minutes.

\begin{table}
	\caption{Summary of the feedback given by developers}
    \label{tab:feedback}
    \centering
    \begin{tabular}{lrrrr}
        \toprule
		& 
		\textbf{ni} &
		\textbf{np} & 
		\textbf{wo} &
		Total \\

		\midrule
		
		The issue in the description is: \\
		\midrule
		relevant                   & 4 & 5 & 2 & 11 \\
		of medium relevance        & 2 & 4 & 1 &  7 \\
		not important              & 1 & 1 & 1 &  3 \\
		not really a testing issue &   & 1 & 1 &  2 \\
		\midrule

		The suggestion provided: \\
		\midrule
		points to the exact solution & 3 &   & 1 &  4 \\
		provides helpful information & 4 & 7 & 3 & 14 \\  
		is not really helpful        & 1 & 2 &   &  3 \\
		is misleading                &   & 1 & 1 &  2 \\
		\midrule

		Developers solve the issue by: \\
		\midrule
		adding a new assertion         & 1 & 4 & 1 &  6 \\ 
		slightly modifying a test case & 1 &   &   &  1 \\
		creating a new test case       & 3 & 1 & 1 &  5 \\
		performing other actions       & 3 & 5 & 3 & 11 \\
		\bottomrule

	\end{tabular}
\end{table}

Developers considered that six transformations can be solved with the addition of an assertion, only one by slightly modifying an existing test case and five with the creation of a new test case. In 11 cases developers considered that the transformations should be solved by other actions. These actions included, for example, a complete replacement of the assertions in a test case, or a combination of new input and new assertions, or, instead of adding a new verification to an existing test case, developers would prefer the create a new test case by repeating the same code with the new assertion. The solutions are in fact influenced by the testing practices of each developer.

Even when developers would not modify the test code as suggested, the report provides information that it generally considered as helpful and the proposed solution as a valid starting point to ease the understanding of the testing issue.

The developers found that the synthesized suggestion was not helpful in three cases, two \textbf{no-propagation} and one \textbf{no-infection}. In two of those cases, the developers wished that the suggestion contained more information about the state differences between the original and the transformed method. For example, developers would like the tool to identify the instruction in the test code that triggered the method invocation for which the state difference was observed. The third case was a method related to the performance of the code. The developer argued that a test case for this method would be ``artificial''.

Two suggestions were considered as misleading. In one case, the issue could not be reproduced by the developer. In the other, the developer could not make the connection between the program state difference and the test cases executing the method. 
So, developers find the suggestions not helpful or misleading when they fail to understand what caused the program state difference between the executions of the original and the transformed method. Suggestions could be further improved with additional information, for example, the stack trace containing the invocation sequence from the test code to method in \textbf{no-propagation} symptoms.

\subsubsection{Actual solutions given by the developers}

Developers actually solved 13 issues with 11 commits. The details of the commits are available online \footnote{https://github.com/STAMP-project/descartes-amplification-experiments/blob/master/validation/commits.md}. One issue in \texttt{funcon4j} was not solved as it was related to a testability problem. No issue was fixed for \texttt{greycat} due to non-technical and unrelated reasons. In this section we provide two examples of how developers solved the extreme transformations and how the solutions relate to the synthesized suggestion.

In project \texttt{funcon4j} the developer was presented with a \textbf{no-infection} symptom. The suggestion, was to create a variant of the existing test case to make the method produce a different value. In the questionnaire, the developer answered that the issue can be solved with the addition of a new assertion.  \autoref{list:fixture:funcon4j} shows the  actual test modification: the addition of line \ref{line:fixture:funcon4j:fix}. The commit can be seen at \url{https://github.com/manuelleduc/Funcon4J/commit/63722262313fb2dac5b516bbae5f04e0502e7f26}.

The developer added a new assertion but she also included a new input derived from the existing code, using a small modification. The actual test improvement corresponds to the suggestion generated by \tool. 

\begin{lstlisting}[caption=Example of a test fixture created by a developer based on the information contained in the \tool report, captionpos=b, label=list:fixture:funcon4j]
test("{a = 1, b = 2} > {b = 1, a = 1};;","true");
test("{a = 1, b = 1} > {b = 1, a = 1};;","false"); // Fixture %*\label{line:fixture:funcon4j:fix}*)
test("\"abc\" > \"abd\";;","false");
\end{lstlisting}

In \autoref{list:fixture:sat4j} we show an example of a test code fixture created by a developer of \texttt{sat4j-core}, with the help of \tool.
The original test code is in lines \ref{line:fixture:sat4j:orig} and \ref{line:fixture:sat4j:orig-end}. The report, available at \url{https://github.com/STAMP-project/descartes-amplification-experiments/blob/master/validation/sat4j-core/selected-issues/6.md} pointed at line \ref{line:fixture:sat4j:issue}. The extreme transformation made the method invocation produce a null value, while the original code should produce a non-null array. The test fixture can be seen between lines \ref{line:fixture:sat4j:fix} and \ref{line:fixture:sat4j:fixend}. The developer confirmed that \tool provided the exact solution and added the proposed assertion (cf. \ref{line:fixture:sat4j:assert}). An extra assertion was also added in the following line.

\begin{lstlisting}[caption=Example of the addition of an assertion guided by the generated report, captionpos=b, label=list:fixture:sat4j]
// Original code
clause.push(-3); %*\label{line:fixture:sat4j:orig}*)
solver.addClause(clause);
int counter = 0;
while (solver.isSatisfiable() && counter < 10) {
  solver.model();  %*\label{line:fixture:sat4j:issue}*)
  counter++;
} %*\label{line:fixture:sat4j:orig-end}*)

// Fixture created by the developer
clause.push(-3); %*\label{line:fixture:sat4j:fix}*)
solver.addClause(clause);
int counter = 0;
int[] model;
while (solver.isSatisfiable() && counter < 10) {
 solver.model();
 model = solver.model();
 assertNotNull(model); //Fixture %*\label{line:fixture:sat4j:assert}*)
 assertEquals(3, model.length); //Fixture %*\label{line:fixture:sat4j:assert2}*)
 counter++;
} %*\label{line:fixture:sat4j:fixend}*)
\end{lstlisting}

\takeaway{rqeval}{Our empirical evaluation shows that the synthesized suggestions are helpful for developers. In the best case, \tool even points to the exact solution. 
The real test fixtures created by the developers to fix the testing problem are in line with the suggestions synthesized by \tool.}

\subsection{\ref{rqsolve}: \rqsolvetext{}}

With this question we explore if state of the art test generation tools can help developers to deal with undetected extreme transformations, \ie they can generate tests that detect the extreme transformations missed by the original test suite.

With current tools, we have two possibilities:  generate test cases from scratch targeting the methods in question; improve existing test cases that are known to reach the method in the extreme transformation. The expected benefit of the first possibility is that test cases are as close as possible to the method. This increases the chances to observe the state differences. Also, some developers prefer to preserve existing test cases (cf.\ref{rqeval:answer}). The second option, uses existing test cases as seeds for the generation process. These tests already have inputs able to execute the method in the transformation. These test cases may just need small code adjustments to make the method return a different value or to actually verify the effects of the transformation. 


Based on the aforementioned options, we implemented two improvement strategies. One strategy is based on \evosuite \cite{fraser_achieving_2015}, a state of the art tool for automatic test generation in Java. It generates test cases from scratch and uses coverage and weak mutation to assess the generated test suites. The other strategy is based on \dspot \cite{danglot_automatic_2019}, is a test improvement tool that can use the ratio of undetected extreme transformation as the objective function. 

We applied both strategies to the projects listed in \autoref{tab:projects}. Since both tools produce non-deterministic results, we attempted the improvement process for each project and strategy \attempts\ times. 

We consider an undetected extreme transformation to be \emph{solved} by one strategy if it is detected by the test cases generated in any of the improvement attempts. In our experiments we evaluate both strategies in terms of the stability of the results, that is, if they obtain similar results in all attempts. We also analyze the absolute number of transformations solved by each strategy in total and considering the type of symptom.

In this section, we briefly summarize how each strategy selects the input and configures the tools. Then we discuss the results we obtained and their implications.

\subsubsection{Detecting extreme transformations with \evosuite}

\evosuite is a test generation tool for \java classes. It implements a search-based approach to produce test suites from scratch. As a fitness function, \evosuite maximizes a combination of coverage criteria and weak mutation score. \evosuite also minimizes the generated test suite to keep only test cases that are valuable according to the fitness function. The tool adds assertions to the test cases based on the observation of the test executions.

In our experiment we identified a set of methods for each project as targets for \evosuite. The set was computed as follows:
\begin{enumerate*}
    \item We include all methods with at least one undetected extreme transformation
    \item If there is a private method in the set, we remove it and add all the accessible methods in the project that could be used instead. This is done in the same way as in stages 2 and 3 of \tool.
    \item If there is a method declared in an abstract class or an interface (default interface methods in \java 8+), then we remove the method from the targets and add all its implementations instead.
   \item The two previous steps are repeated until no further change can be done to the set.
\end{enumerate*}

The second column in \autoref{tab:targets} shows the number of target methods identified for each study subject. Recall that the goal is to create a test cases specifying the methods with undetected extreme transformations, using these identified target methods as entry points.

Next, we provide the target methods as inputs for \evosuite.  We use the \evosuite default parameter configuration. The outcome, for a specific project, is a set of test cases generated for all the target methods in that project.
As final step, we run the generated test cases to determine if the \evosuite tests can detect the extreme transformations that were previously undetected.

\subsubsection{Detecting extreme transformations with \dspot}

\dspot is a tool designed to improve test cases written by developers. It takes as input existing test cases and gradually improves their fault detection capabilities. \dspot transforms the input of the test cases by altering literals or adding and removing method calls. Then, new regression assertions are inserted to the modified test cases. The assertions are built using the values observed with existing getter methods. \dspot only keeps the generated test cases that increment the mutation score of the existing test suite. As a proxy for the mutation score, we used the ratio of detected extreme transformations in the code covered by the test to improve.

It is prohibitively expensive to use \dspot and improve all test cases executing one method and do the same for every method with undetected extreme transformations. Recall that some of these methods are executed by more than 100 test cases (cf. \autoref{tab:projects}). For this reason, we instruct \dspot to improve the test case that is the closest, in a sequence call to a method where there is an undetected extreme transformation. 

We detect the closest test case by executing the test suite and obtaining the stack trace from any method in a test class to the application methods with undetected extreme transformations. For each application method, the closest test case is the test method that produces the shortest stack trace. Different methods may be executed by the same test cases. In some situations, the closest test case to more than one method is the same. In such a situation we keep the same test case for these methods. 

The selection results in a set of test cases for each project. The third column in \autoref{tab:targets} shows the number of test cases identified for each study subject. 

An attempt to solve the undetected extreme transformations for a given project, consists in executing \dspot for each test case in the identified set. We used all \dspot search operators (amplifiers) available (by default \dspot only adds regression assertions to existing test cases) and performed three amplification iterations for each target. The output of one attempt is the combination of all the improved test cases.

\begin{table}
    \caption{Targets identified for automatic test generation and test improvement. The \textit{Target methods} column shows the methods identified as targets to be used with \evosuite. The \textit{Target test cases} are the test cases selected for improvement with \dspot}
    \label{tab:targets}
    \centering
    \begin{tabular}{lrr}
        \toprule
        Project               & Target methods & Target test cases \\
        \midrule
        jpush-api-java-client &              2 &                 2 \\
        commons-cli           &              6 &                 3 \\
        jopt-simple           &              2 &                 3 \\
        yahoofinance-api      &             11 &                 1 \\
        gson-fire             &              7 &                 3 \\
        j2html                &              3 &                 2 \\
        spring-petclinic      &              6 &                 5 \\
        javapoet              &             10 &                10 \\
        eaxy                  &             21 &                 8 \\
        java-html-sanitizer   &             13 &                11 \\
        cron-utils            &             28 &                10 \\
        commons-codec         &             30 &                18 \\
        jsoup                 &             60 &                26 \\
        TridentSDK            &             49 &                13 \\
        jcodemodel            &            147 &                25 \\
        \midrule
        Total                 &            395 &               140 \\
        \bottomrule
    \end{tabular}
\end{table}

\subsubsection{Experimental Results}

\begin{figure}
    \centering
    \includegraphics[width=2.5in]{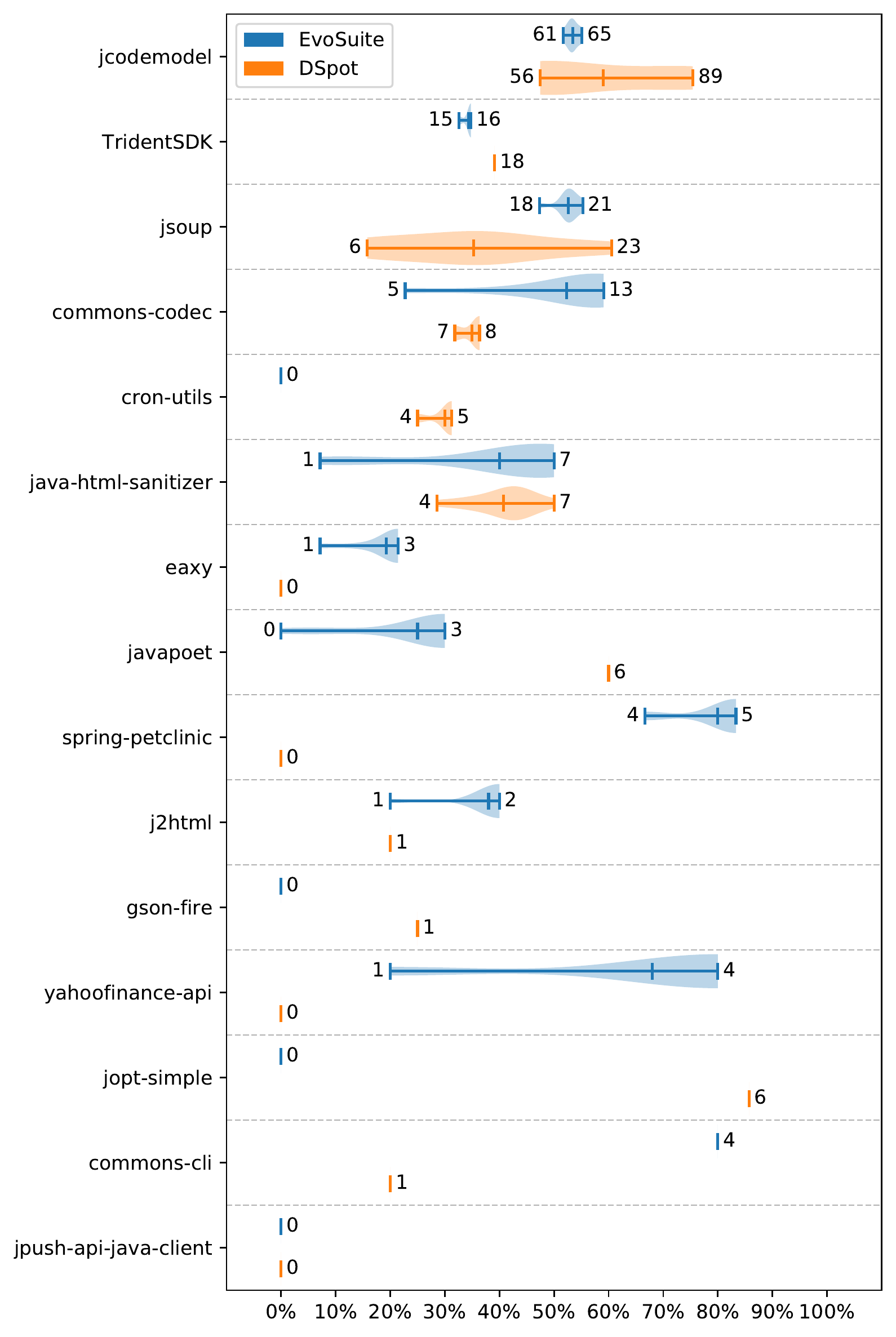}
    \caption{Automated test improvement for 15 projects}
    \label{fig:improvement}
\end{figure}

\autoref{fig:improvement} shows the distribution of the relative improvement obtained for each project. The x-axis is the percentage of undetected extreme transformations that the new test cases can detect.The figure also shows the number of extreme transformations detected for the worst and best attempts. 

Both strategies  produce test improvement for 11/15 projects. Eight projects are improved  by both strategies. No strategy can generate tests that fix undetected extreme transformations in  \texttt{jpush-api-java-client}.  \dspot produces the best improvement in one attempt for seven projects while  \evosuite  achieves the best improvement for six projects. The same best improvement is obtained for \texttt{java-html-sanitizer} and  \texttt{jpush-aoi-java-client}. 

No strategy reaches an improvement beyond 85\% in one attempt.  \dspot  reaches this percentage by handling 6/7 extreme transformations in \texttt{jopt-simple}. The \evosuite strategy achieves an 83\% improvement handling 5/6 extreme transformations in \texttt{spring-petclinic}. In the largest project, \texttt{jcodemodel}, the \evosuite strategy handles 64/118 transformations while \dspot handles 89/118. The improvement is below 75\% for both cases.

This means that no strategy was considerably better than the other in terms of handling all projects and even all the transformations on each project.

Inspecting the improvement distributions we notice that \dspot has better convergence in all projects but \texttt{jcodemodel} and \texttt{jsoup}. This is a consequence of using the existing test cases as seeds for the generation process. Those test cases have been created by developers, therefore, they are close to a local optimum and \dspot, as it implements a local search approach is able to converge more frequently to this solution.

We computed the set of solved extreme transformations for each strategy. That is, the extreme transformations detected in at least one attempt in one project by each strategy. \evosuite solves \essol\ (\percentage\essol\tuet) of all \tuet\ transformations. On its side, the \dspot strategy solved \dssol\ (\percentage\dssol\tuet). Only \esdsin\ (\percentage\esdsin\tuet) transformations are solved by both strategies. A total of \esdsun\ (\percentage\esdsun\tuet) are solved by any of the strategies, which leaves \unsolved\ (\percentage\unsolved\tuet) transformations unsolved. 

\dspot   solves more extreme transformations, but no strategy solves more than 60\% of all transformations. The intersection between the transformations solved by both strategies is low, below 30\%. Only when both results are combined, we are able to solve \percentage\esdsun\tuet\ of all transformations but still remains below 80\%. This is a direct result of the very diverse nature of undetected extreme transformations. A subset of the transformations is more challenging for one of the strategies than the other.

\autoref{tab:symptom-improvement} shows the set of solved extreme transformations by symptom (It is only a coincidence that the intersection for all symptoms is 29). \dspot performs better for all symptoms but more notable for \textbf{weak-oracle} symptoms. These are precisely the symptoms that \tool identifies that benefit the most from modifying existing test cases, in particular, the addition of new assertions, which is one of the main features of \dspot.

\begin{table}
    \caption{Solved transformations according to their symptom. Second and third columns show the transformations solved by \evosuite and \dspot respectively. \emph{Intersection} shows the number of transformations solved by both tools and \emph{Union} shows those transformations solved by at least one strategy.}
    \label{tab:symptom-improvement}
    \centering
    \begin{tabular}{lrrrr}
        \toprule
        Symptom                 & \evosuite &  \dspot & Intersection &      Union \\ 
        \midrule
        \textbf{no-infection}   &        53 &      57 &           29 &    \solni\ \\
        \textbf{no-propagation} &        55 &      66 &           29 &    \solnp\ \\
        \textbf{weak-oracle}    &        38 &      57 &           29 &    \solwo\ \\
        \midrule
        Total                   &   \essol\ & \dssol\ &     \esdsin\ &   \esdsun\ \\
        \bottomrule
    \end{tabular}
\end{table}

\dspot  produces better results because it uses the test cases written by developers as seed. 
We believe that this strategy is closer to what a developer would do. The developer actions exemplified in \ref{rqeval:answer} provide supporting evidence in this sense. However, the strategy based on \evosuite does not fall too much behind. A better fine-tuning of the tool and the incorporation of a notion extreme transformation to the fitness function may help the strategy achieve a much better result.


This opens the opportunity for a more targeted solution that fully exploits the results of \tool. Nevertheless, we observe in \ref{rqeval:answer} that developers have strong and diverse opinions when it comes to refactor the testing code to deal with extreme transformations. Providing them with helpful and well localized information extracted from dynamic and static analysis and perhaps small code changes might be the adequate solution instead of a fully automated approach.

\takeaway{rqsolve}{
    \dspot olves \percentage\dssol\tuet\ of the undetected transformations, while the \evosuite solves \percentage\essol\tuet\ . None of state of the art test generation tools can fix all testing issues revealed by extreme transformations. 
}

\section{Threats to validity}

As said before, we believe that the definitions included in \autoref{sec:definitions} and the entire process implemented in \tool can be extrapolated to other programming languages and runtimes. However we can not assure that the results we have obtained for \ref{rqinfect} and \ref{rqpropagate} may generalize to other environments and even a different set of \java projects. The set of study subjects used to answer these two questions include only 15 projects, which is not large enough to derive statistically robust results. This selection is also restricted to small to medium sized \maven projects with a single module. In large and multimodule projects the situation may be different. Even though, our set of subjects is diverse with regards to the obtained results.

The tooling we have developed and also the external tools we have used are not exempt of bugs. \tool has been conceived as an open-source project available in \github. We have used \pit 1.4.7, and \descartes 1.2.5 to answer all questions. We used  \dspot 2.1.0 and built \evosuite from revision 1895b6d to answer \ref{rqsolve}. Using a different version of these tools may produce a different result given their natural evolution.

The validation to answer \ref{rqeval} was conducted over a very small set of projects and issues. Also, the selection included projects whose developers we know and can reach. In order to be able to generalize these results a more impartial and larger evaluation is required. However, we selected projects with different application domains and developers with different backgrounds and experience.

\section{Related Work}


In 1978, DeMillo and colleagues proposed a seminal piece of work in the area of software testing; ``Hints on Test Data Selection: Help for the Practicing Programmer''  \cite{demillo_hints_1978}. This work is mostly known for introducing mutation analysis as a novel test assessment criterion, stronger than code coverage, and for the whole literature that has followed. 
Meanwhile, our work is more inspired by the paper's intention to  provide ``hints on test data selection''.
In particular, the following concluding remark is a key motivation for our work: mutation analysis ``yields advice'' to be used in generating test data for similar programs''. 
While this quote dates from four decades back, there has been very little work that pursued this idea of ``advice'' extraction from mutation analysis. Our work explores this opportunity further, in the context of extreme transformations.



Extreme transformations, originally named extreme mutations, can be seen as a ``lightweight'' form of mutation testing \cite{niedermayr_will_2016, niedermayr_is_2019}. introduced by Niedermayr and colleagues \cite{niedermayr_will_2016} to discover \textit{pseudo-tested methods}, that is, those for which no extreme transformation is detected by the test suite. 
Recently,  Niedermayr and Wagner \cite{niedermayr_is_2019} showed a correlation between the stack distance from the test code to the pseudo-tested method and the test effectiveness. A part of our present work supports this observation with evidence that the effects of methods with higher stack distance are more likely to be masked in the invocation sequence. The core of our contribution is completely novel: automatically analyze undetected extreme transformations to generate suggestions to be used in improving the test suite.

\subsection{Reachability, Infection, Propagation}

The \emph{Reachability, Infection, Propagation} model (RIP) \cite{morell_theory_1984,morell_theory_1990,demillo_constraint-based_1991} states the necessary conditions for a fault to be detected. That is:
\begin{enumerate*}
    \item the fault should be reached/executed
    \item the program state should be changed/infected after the fault is executed
    \item the program infection should propagate to the output.
\end{enumerate*}
Our work is inspired by the RIP model and considers that the propagated infection should be \emph{asserted} to be discovered in automatic unit testing. We use extreme transformations as fault surrogates.

Voas introduced the \emph{propagation}, \emph{infection} and \emph{execution} (\emph{PIE}) analysis \cite{voas_pie:_1992}. 
Subsequently, Voas and Miller \cite{voas_software_1995} discussed the notion of \emph{testability} in relation to the PIE analysis: ``\textit{the likelihood that the code can fail if something in the code is incorrect}''. The authors express concerns about the loss of observation ability linked to information hiding in object-oriented programs.  Our work contributes to the state of the art in this area with novel results about infections that do not propagate to the test cases. 

Li and Offutt \cite{li_test_2017} evolve the RIP model into \emph{Reachability Infection Propagation Revealability}, (\emph{RIPR}) to include the notion that the oracle in automated tests must \emph{reveal} failures. They devise several  rules to specify which program states to check. The  \textbf{assert} condition we present is a parallel to their  notion of \emph{Revealability} .
Lin \etal{} \cite{lin_domain-rip_2019} recently proposed D-RIP, a combination of RIP analysis with domain-based testing. They conclude that a program infection that does not propagate is the main reason why a mutant is not detected. This coincides with our own observations in the present study.

Since its introduction, the RIP model has been leveraged to analyze faults and the conditions under which they can be detected. Yet, the creation of actionable suggestions for developers has received little attention.

\subsection{Mutation-based test generation}

DeMillo and Offutt \cite{demillo_constraint-based_1991} generate test data based on mutation testing. They derive the conditions/constraints under which the execution of a mutant produces a different state than the original code and then generate test inputs with a constraint solver. 
Later works relied on search-based techniques to generate test cases that target live mutants, using the mutation score as a fitness function.
In earlier work, we introduced a version of genetic algorithm to generate test cases in C\# that could kill live mutants \cite{Baudry05d}. 
Fraser and Zeller \cite{fraser_mutation-driven_2012} observe the objects during the test generation process and use the observed values generate assertions that can distinguish  the original code apart from its mutants. 
The idea is further explored by Fraser and Arcuri and implemented in \evosuite \cite{fraser_achieving_2015}. 

These works use traditional mutation testing to guide automatic test generation.  Yet, none of these works validate the relevance of mutants for developers nor the utility of the generated tests. In our work we leverage extreme mutation (transformations) with the explicit goal to provide guidance to developers. Our suggestions seek to improve the test suite in terms of fault detection while letting control to the developers about how they want to use their test resources..

\subsection{Test improvement}

Beyond extreme transformations, our work relates to the broader field of works that aim at assisting developers to build stronger test suites  \cite{danglot_snowballing_2017}. Here we discuss the most related works that target the improvement of test oracles.

Staats \etal{} \cite{staats_automated_2012} propose an automated process to support the creation of test oracles defined as concrete values the system under test is expected to produce. Their goal is to guide the tester in the selection of variables and outputs for which the values should be specified. They leverage mutation analysis to rank variables and outputs with respect to their ability at  revealing a mutant. In the current work we suggest developers ways to enhance the assertion-based oracles. In particular we provide the expected values and methods that can be used to observe these values.

Daniel \etal developed ReAssert \cite{daniel_reassert_2009} a tool that suggests test repairs to developers. The tool tries to change the behavior of failing test cases to make them pass while trying to maintain the original test code as much as possible and retaining the test's regression detection capabilities.  \tool, like ReAssert, focuses on improving the test code. ReAssert instruments the test code to observe expected and actual values in assertions. We instrument the program and the test code to observe the program state at different levels. ReAssert applies predefined repair strategies to the tests. Our suggestions are built solely form the observations.

Xie \cite{xie_augmenting_2006} introduces Orstra, a tool to automate the creation of regression oracles. 
Pacheco and Ernst have developed Randoop \cite{pacheco_randoop_2007} and Eclat \cite{pacheco_eclat_2005}, based on their experience with Daikon \cite{ernst_dynamically_2001} to monitor runtime values and infer valuable properties to build an oracle a tool for automatic test generation. T
All these works leverage the observation during test executions to create regression oracles and new assertions. \tool works under the same principles. It observes the program state at different levels when executing the tests. The observed values become oracles to check the values produced in the execution of extreme transformations. A difference is converted into a suggestion for the developers.

Song and colleagues \cite{song_unitplus_2007} have a that tool recommends observer methods that could be used to verify the effects of methods that change the state of the receiver. Their analysis is similar to ours when we select methods to fix \textbf{weak-oracle} symptoms.

\section{Conclusions}

In this work we describe an infection-propagation analysis to generate actionable suggestions that can help developers deal with undetected extreme transformations. The process is implemented in an open-source tool, \tool, that can target \java projects built with \maven.
With the help of \tool we study the undetected extreme transformations of 15 \java open-source projects. In \tuet\ transformations:
\begin{itemize}
    \item \percentage\tni\tuet\ do not infect the immediate program state
    \item \percentage\tnp\tuet\ infect the immediate program state but the infection do not propagate to an observable point
    \item \percentage\twor\tuet\ propagate the infection to an observable point, yet the existing test cases do not assess the program state items that are affected.    
\end{itemize}
We validated the suggestions generated by \tool with the help of the developers of three open-source projects. Most suggestions provided helpful information or the exact solution to detect the extreme transformations.
We also explored two automatic strategies to solve undetected extreme transformations. These strategies were based in state-of-the-art test generation and test improvement tools. The \dspot based strategy produces better results as it uses the ratio of  undetected extreme transformations as fitness function and a seed that already executes the method involved in the transformation.
Since undetected extreme transformations point to very well localized testing issues, more targeted automatic solutions are to be explored and compared in the future.

%% file: main.bbl
\begin{thebibliography}{10}
\providecommand{\url}[1]{#1}
\csname url@samestyle\endcsname
\providecommand{\newblock}{\relax}
\providecommand{\bibinfo}[2]{#2}
\providecommand{\BIBentrySTDinterwordspacing}{\spaceskip=0pt\relax}
\providecommand{\BIBentryALTinterwordstretchfactor}{4}
\providecommand{\BIBentryALTinterwordspacing}{\spaceskip=\fontdimen2\font plus
\BIBentryALTinterwordstretchfactor\fontdimen3\font minus
  \fontdimen4\font\relax}
\providecommand{\BIBforeignlanguage}[2]{{%
\expandafter\ifx\csname l@#1\endcsname\relax
\typeout{** WARNING: IEEEtran.bst: No hyphenation pattern has been}%
\typeout{** loaded for the language `#1'. Using the pattern for}%
\typeout{** the default language instead.}%
\else
\language=\csname l@#1\endcsname
\fi
#2}}
\providecommand{\BIBdecl}{\relax}
\BIBdecl

\bibitem{niedermayr_will_2016}
R.~Niedermayr, E.~Juergens, and S.~Wagner, ``\BIBforeignlanguage{en}{Will my
  tests tell me if {I} break this code?}'' in
  \emph{\BIBforeignlanguage{en}{Proceedings of the {International} {Workshop}
  on {Continuous} {Software} {Evolution} and {Delivery}}}.\hskip 1em plus 0.5em
  minus 0.4em\relax ACM Press, 2016, pp. 23--29.

\bibitem{vera_comprehensive_2018}
\BIBentryALTinterwordspacing
O.~L. Vera-P{\'e}rez, B.~Danglot, M.~Monperrus, and B.~Baudry, ``A
  comprehensive study of pseudo-tested methods,'' \emph{Empirical Software
  Engineering}, Sep 2018. [Online]. Available:
  \url{https://doi.org/10.1007/s10664-018-9653-2}
\BIBentrySTDinterwordspacing

\bibitem{morell_theory_1990}
L.~J. Morell, ``A theory of fault-based testing,'' \emph{IEEE Transactions on
  Software Engineering}, vol.~16, no.~8, pp. 844--857, Aug. 1990.

\bibitem{demillo_constraint-based_1991}
R.~A. DeMillo and A.~J. Offutt, ``Constraint-based automatic test data
  generation,'' \emph{IEEE Transactions on Software Engineering}, vol.~17,
  no.~9, pp. 900--910, Sep. 1991.

\bibitem{voas_pie:_1992}
J.~M. Voas, ``{PIE}: a dynamic failure-based technique,'' \emph{IEEE
  Transactions on Software Engineering}, vol.~18, no.~8, pp. 717--727, Aug.
  1992.

\bibitem{danglot_automatic_2019}
\BIBentryALTinterwordspacing
B.~Danglot, O.~L. Vera-Pérez, B.~Baudry, and M.~Monperrus,
  ``\BIBforeignlanguage{en}{Automatic test improvement with {DSpot}: a study
  with ten mature open-source projects},''
  \emph{\BIBforeignlanguage{en}{Empirical Software Engineering}}, vol.~24,
  no.~4, pp. 2603--2635, Aug. 2019. [Online]. Available:
  \url{https://doi.org/10.1007/s10664-019-09692-y}
\BIBentrySTDinterwordspacing

\bibitem{chiba_load-time_2000}
S.~Chiba, ``\BIBforeignlanguage{en}{Load-{Time} {Structural} {Reflection} in
  {Java}},'' in \emph{\BIBforeignlanguage{en}{{ECOOP} 2000 —
  {Object}-{Oriented} {Programming}}}, ser. Lecture {Notes} in {Computer}
  {Science}, E.~Bertino, Ed.\hskip 1em plus 0.5em minus 0.4em\relax Springer
  Berlin Heidelberg, 2000, pp. 313--336.

\bibitem{pawlak_spoon_2016}
R.~Pawlak, M.~Monperrus, N.~Petitprez, C.~Noguera, and L.~Seinturier,
  ``{SPOON}: {A} library for implementing analyses and transformations of
  {Java} source code,'' \emph{Softw., Pract. Exper.}, vol.~46, pp. 1155--1179,
  2016.

\bibitem{vera_descartes_2018}
\BIBentryALTinterwordspacing
O.~L. Vera-P\'erez, M.~Monperrus, and B.~Baudry, ``Descartes: A {PITest} engine
  to detect pseudo-tested methods,'' in \emph{Proceedings of the 2018 33rd
  ACM/IEEE International Conference on Automated Software Engineering (ASE
  ’18)}, 2018, pp. 908--911. [Online]. Available:
  \url{https://dl.acm.org/citation.cfm?doid=3238147.3240474}
\BIBentrySTDinterwordspacing

\bibitem{urli_how_2018}
\BIBentryALTinterwordspacing
S.~Urli, Z.~Yu, L.~Seinturier, and M.~Monperrus, ``How to design a program
  repair bot?: Insights from the repairnator project,'' in \emph{Proceedings of
  the 40th International Conference on Software Engineering: Software
  Engineering in Practice}, ser. ICSE-SEIP '18.\hskip 1em plus 0.5em minus
  0.4em\relax New York, NY, USA: ACM, 2018, pp. 95--104. [Online]. Available:
  \url{http://doi.acm.org/10.1145/3183519.3183540}
\BIBentrySTDinterwordspacing

\bibitem{fraser_achieving_2015}
G.~Fraser and A.~Arcuri, ``\BIBforeignlanguage{en}{Achieving scalable
  mutation-based generation of whole test suites},''
  \emph{\BIBforeignlanguage{en}{Empirical Software Engineering}}, vol.~20,
  no.~3, pp. 783--812, Jun. 2015.

\bibitem{demillo_hints_1978}
R.~A. DeMillo, R.~J. Lipton, and F.~G. Sayward, ``Hints on {Test} {Data}
  {Selection}: {Help} for the {Practicing} {Programmer},'' \emph{Computer
  Magazine}, vol.~11, no.~4, pp. 34--41, Apr. 1978.

\bibitem{niedermayr_is_2019}
\BIBentryALTinterwordspacing
R.~Niedermayr and S.~Wagner, ``Is the {Stack} {Distance} {Between} {Test}
  {Case} and {Method} {Correlated} {With} {Test} {Effectiveness}?'' in
  \emph{Proceedings of the {Evaluation} and {Assessment} on {Software}
  {Engineering}}, ser. {EASE} '19.\hskip 1em plus 0.5em minus 0.4em\relax New
  York, NY, USA: ACM, 2019, pp. 189--198, event-place: Copenhagen, Denmark.
  [Online]. Available: \url{http://doi.acm.org/10.1145/3319008.3319021}
\BIBentrySTDinterwordspacing

\bibitem{morell_theory_1984}
\BIBentryALTinterwordspacing
L.~J. Morell, ``\BIBforeignlanguage{en}{A {Theory} of {Error}-{Based}
  {Testing}.}'' MARYLAND UNIV COLLEGE PARK DEPT OF COMPUTER SCIENCE, Tech. Rep.
  TR-1395, Apr. 1984. [Online]. Available:
  \url{https://apps.dtic.mil/docs/citations/ADA143533}
\BIBentrySTDinterwordspacing

\bibitem{voas_software_1995}
J.~M. Voas and K.~W. Miller, ``Software testability: the new verification,''
  \emph{IEEE Software}, vol.~12, no.~3, pp. 17--28, May 1995.

\bibitem{li_test_2017}
N.~Li and J.~Offutt, ``Test {Oracle} {Strategies} for {Model}-{Based}
  {Testing},'' \emph{IEEE Transactions on Software Engineering}, vol.~43,
  no.~4, pp. 372--395, Apr. 2017.

\bibitem{lin_domain-rip_2019}
H.~Lin, Y.~Wang, Y.~Gong, and D.~Jin, ``Domain-{RIP} {Analysis}: {A}
  {Technique} for {Analyzing} {Mutation} {Stubbornness},'' \emph{IEEE Access},
  vol.~7, pp. 4006--4023, 2019.

\bibitem{Baudry05d}
\BIBentryALTinterwordspacing
B.~Baudry, F.~Fleurey, J.-M. J{'e}z{'e}quel, and Y.~{Le~Traon}, ``Automatic
  test cases optimization: a bacteriologic algorithm,'' \emph{IEEE Software},
  vol.~22, no.~2, pp. 76--82, Mar. 2005. [Online]. Available:
  \url{http://www.irisa.fr/triskell/publis/2005/Baudry05d.pdf}
\BIBentrySTDinterwordspacing

\bibitem{fraser_mutation-driven_2012}
G.~Fraser and A.~Zeller, ``Mutation-{Driven} {Generation} of {Unit} {Tests} and
  {Oracles},'' \emph{IEEE Transactions on Software Engineering}, vol.~38,
  no.~2, pp. 278--292, Mar. 2012.

\bibitem{danglot_snowballing_2017}
\BIBentryALTinterwordspacing
B.~Danglot, O.~Vera{-}Perez, Z.~Yu, M.~Monperrus, and B.~Baudry, ``A
  snowballing literature study on test amplification,'' \emph{arXiv preprint:
  1705.10692}, vol. abs/1705.10692, 2017. [Online]. Available:
  \url{http://arxiv.org/abs/1705.10692}
\BIBentrySTDinterwordspacing

\bibitem{staats_automated_2012}
M.~Staats, G.~Gay, and M.~P.~E. Heimdahl, ``Automated oracle creation support,
  or: {How} {I} learned to stop worrying about fault propagation and love
  mutation testing,'' in \emph{2012 34th {International} {Conference} on
  {Software} {Engineering} ({ICSE})}, Jun. 2012, pp. 870--880.

\bibitem{daniel_reassert_2009}
\BIBentryALTinterwordspacing
B.~Daniel, V.~Jagannath, D.~Dig, and D.~Marinov, ``{ReAssert}: {Suggesting}
  {Repairs} for {Broken} {Unit} {Tests},'' in \emph{Proceedings of the 2009
  {IEEE}/{ACM} {International} {Conference} on {Automated} {Software}
  {Engineering}}, ser. {ASE} '09.\hskip 1em plus 0.5em minus 0.4em\relax
  Washington, DC, USA: IEEE Computer Society, 2009, pp. 433--444. [Online].
  Available: \url{https://doi.org/10.1109/ASE.2009.17}
\BIBentrySTDinterwordspacing

\bibitem{xie_augmenting_2006}
T.~Xie, ``Augmenting {Automatically} {Generated} {Unit}-{Test} {Suites} with
  {Regression} {Oracle} {Checking},'' in \emph{{ECOOP} 2006 –
  {Object}-{Oriented} {Programming}}, ser. Lecture {Notes} in {Computer}
  {Science}.\hskip 1em plus 0.5em minus 0.4em\relax Springer, Berlin,
  Heidelberg, Jul. 2006, pp. 380--403.

\bibitem{pacheco_randoop_2007}
C.~Pacheco and M.~D. Ernst, ``Randoop: {Feedback}-directed {Random} {Testing}
  for {Java},'' in \emph{Companion to the 22Nd {ACM} {SIGPLAN} {Conference} on
  {Object}-oriented {Programming} {Systems} and {Applications} {Companion}},
  ser. {OOPSLA} '07.\hskip 1em plus 0.5em minus 0.4em\relax New York, NY, USA:
  ACM, 2007, pp. 815--816.

\bibitem{pacheco_eclat_2005}
\BIBentryALTinterwordspacing
------, ``\BIBforeignlanguage{en}{Eclat: {Automatic} {Generation} and
  {Classification} of {Test} {Inputs}},'' in
  \emph{\BIBforeignlanguage{en}{{ECOOP} 2005 - {Object}-{Oriented}
  {Programming}}}, ser. Lecture {Notes} in {Computer} {Science}.\hskip 1em plus
  0.5em minus 0.4em\relax Springer, Berlin, Heidelberg, Jul. 2005, pp.
  504--527. [Online]. Available:
  \url{https://link.springer.com/chapter/10.1007/11531142_22}
\BIBentrySTDinterwordspacing

\bibitem{ernst_dynamically_2001}
M.~D. Ernst, J.~Cockrell, W.~G. Griswold, and D.~Notkin, ``Dynamically
  discovering likely program invariants to support program evolution,''
  \emph{IEEE Transactions on Software Engineering}, vol.~27, no.~2, pp.
  99--123, Feb. 2001.

\bibitem{song_unitplus_2007}
\BIBentryALTinterwordspacing
Y.~Song, S.~Thummalapenta, and T.~Xie, ``{UnitPlus}: {Assisting} {Developer}
  {Testing} in {Eclipse},'' in \emph{Proceedings of the 2007 {OOPSLA}
  {Workshop} on {Eclipse} {Technology} {eXchange}}, ser. eclipse '07.\hskip 1em
  plus 0.5em minus 0.4em\relax New York, NY, USA: ACM, 2007, pp. 26--30,
  event-place: Montreal, Quebec, Canada. [Online]. Available:
  \url{http://doi.acm.org/10.1145/1328279.1328285}
\BIBentrySTDinterwordspacing

\end{thebibliography}
